\documentclass[reprint,3p,times,onecolumn,12pt]{elsarticle} 
\usepackage{amsmath,amsthm,amscd,color}
\usepackage{amsfonts}
\usepackage{amssymb}
\usepackage{graphicx}
\usepackage{latexsym}
\usepackage{epstopdf}

\usepackage{soul}
\biboptions{sort&compress}
\usepackage{setspace}
\usepackage{xcolor}

\newcommand{\cb}{\color{blue}}

\newcommand{\bea}{\begin{eqnarray}}
\newcommand{\eea}{\end{eqnarray}}
\newcommand{\bes}{\begin{subequations}}
	\newcommand{\ees}{\end{subequations}}

\usepackage[colorlinks=true,bookmarks=false,citecolor=blue,urlcolor=blue]{hyperref}

\usepackage{fancyhdr}
\makeatletter
\def\ps@pprintTitle{%
	\let\@oddhead\@empty
	\let\@evenhead\@empty
	\def\@oddfoot{Ref.: {\cb Physica D: Nonlinear Phenomena {\bf 448} (2023) 133694} \hfill DOI: \href{https://doi.org/10.1016/j.physd.2023.133694}{10.1016/j.physd.2023.133694}}%
	\let\@evenfoot\@oddfoot
}
\makeatother

\begin{document} 
	\title{Bright Matter-Wave Bound Soliton Molecules in Spin-1 Bose-Einstein Condensates with Non-autonomous Nonlinearities}
	
	\author[apctp]{K. Sakkaravarthi \corref{cor}}
	\author[psg]{R. Babu Mareeswaran}
	\author[bhc]{T. Kanna}
	
	\address[apctp]{Young Scientist Training Program, Asia-Pacific Center for Theoretical Physics (APCTP),\newline POSTECH Campus, Pohang - 37673, Republic of Korea} 
	\address[psg]{Department of Physics, PSG College of Arts and Science, Coimbatore - 641014, Tamil Nadu, India} 
	\address[bhc]{Nonlinear Waves Research Lab, PG and Research Department of Physics, Bishop Heber College (Autonomous),\newline Affiliated to Bharathidasan University, Tiruchirappalli - 620017, Tamil Nadu, India} 
	
	\cortext[cor]{
		Email: ksakkaravarthi@gmail.com (K. Sakkaravarthi)\newline 
		\indent \quad Email: babu\_nld@rediffmail.com (R. Babu Mareeswaran)\newline 
		\indent \quad Email: kanna\_phy@gmail.com (T. Kanna)}
	
	\journal{Physica D: Nonlinear Phenomena}
	
	\setstretch{1.213}
	\begin{abstract}
		This work deals with matter-wave bright soliton molecules in spin-1 Bose-Einstein condensates described by three-component Gross-Pitaevskii equations with non-autonomous nonlinearities that can be tuned by Feshbach resonance management. Notably, it portrays the possibility of generating bright bound soliton molecules with the help of an exact analytic solution under a controlled velocity resonance mechanism. Results show that these soliton molecules experience the effects of time-varying nonlinearities and modulate themselves during propagation by keeping their stable properties. Significantly, the chosen periodic and kink-like nonlinearities expose the snaking, bending, compression, and amplification of multi-structured soliton molecules along with appreciable changes in their amplitude, velocity, width, and oscillations of the molecule profiles. The present results will add significant knowledge to a complete understanding beyond the known interaction dynamics of matter-wave bright solitons in spinor condensates.\\~~
		
	\end{abstract}
	
	\maketitle
	
	\noindent{\bf Highlights}
	\begin{itemize}
		\item Investigated the dynamics of bright matter-wave soliton molecules in spinor BECs. 
		\item Constructed explicit solution through similarity transformation and bilinear method.
		\item Analyzed the generation of bright soliton molecule through velocity resonance mechanism.
		\item Explored the effects of periodic and kink-like nonlinearities in bound soliton molecules.
		\item Observed the breathing, amplification, bending, and compression of chain-like soliton molecules.
	\end{itemize} 
	
	\setstretch{1.3}
	\section{Introduction}
	The study of localized nonlinear waves in atomic condensates has become one of the frontier topics in modern-day research. Especially, the  recent experimental discoveries \cite{Pethick-book,PGK-book,NatPRL} have led to a tremendous research interest in Bose-Einstein condensates (BECs). Several works in theoretical and experimental directions on multicomponent BECs demonstrate the existence of different types of nonlinear waves, ranging from bright, gray, dark, and gap solitons to general, Akhmediev and Kuznetsov-Ma breathers, and rogue waves under various homogeneous and controllable environments \cite{Pethick-book}. The realm of multicomponent BECs with optical dipole trap leads to realising the spinor BECs (see review article \cite{PhysRep} and references therein). For a generic hyperfine spin $F = 1$ state, the spinor solitons in a one-dimensional coupled Gross-Pitaevskii (GP) system have been studied theoretically \cite{Ohta,Nat-bec, numeric} and matter rogue waves have been investigated \cite{ref6}. Indeed, pertinent experimental works have shown that the creation of mixed type dark--bright--bright and dark--dark--bright solitons and  magnetic solitons in spin-1 BECs \cite{becexp2,becexp2a,becexp2b,ref7,ref8}. Further, the study of BECs with spatially or temporally varying nonlinearities has emerged as an interesting avenue due to the Feshbach resonance \cite{review2021}, a mechanism by which the scattering lengths can be tuned. Precisely, the spin-dependent interactions can be controlled by employing an external optical or microwave field in the $F=1$ spinor condensate that drives the non-autonomous nonlinearities \cite{ref9,ref10}. By utilizing the above technique, the dynamics of non-autonomous bright matter-wave solitons and interactions depicting soliton amplification with compression have been explored theoretically using the kink-like modulated nonlinearity \cite{tkpla14}. The dynamics of non-autonomous bright-dark solitons and rogue waves have been investigated by both analytical and numerical methods for different temporally modulated nonlinearities \cite{ref12}. Particularly, the dynamics of non-autonomous multi-rogue waves in a three-component GP system is studied and certain interesting features for single, double and triple-hump vector rogue waves are revealed \cite{ref13}. 
	
	Though adequate results on the dynamics of soliton propagation and their interactions are available, their complete behaviour under various controllable physical settings requires further exploration and continues to be an important aspect of the study. To mention a few significant studies along this direction are dark--anti-dark spinor solitons \cite{pgk-pra20,cnsns23}, dark-dark soliton breathing patterns \cite{pgk-jpb21}, soliton interactions on nonzero background \cite{pgk-epjp21}, bright-dark solitons \cite{nld2017, cnsns2022, cnsns2022a}, and several other types of nonlinear waves such as breather and rogue waves \cite{csf2020}, that are obtained through inverse scattering transformation, Hirota bilinear procedure, KP-hierarchy reduction method, Darboux and B{\"a}cklund transformation techniques, etc. 
	In particular, as a branch of soliton interactions, the nature of soliton molecule (SM) generation and their dynamics is of current importance in different nonlinear systems arising from optics to hydrodynamics \cite{mole-boris, mole1}. Soliton molecules are nothing but multiple composite solitons travelling together with resonance velocity (phase matching), which undergo periodic attraction and repulsion with remarkable stability during their propagation. 
	The bound states involving two bright solitons exist when they are brought close to each other but with opposite phases, where the attractive interaction turns repulsive at a particular separation and a stable equilibrium is created and they have been called as soliton molecules due to their ability to restitute their equilibrium separation \cite{pra-mole-2008}. Further, the formation of matter-wave soliton molecules is achieved by a bound pair of two solitons prepared in phase with a sufficiently small separation and relative velocity \cite{njp2011}. 
	One of the striking features is that the soliton molecules propagate around the equilibrium separation due to the available nonlinearity, which helps to minimize/avoid standard crossing interactions. 
	In nonlinear optical systems, it has been achieved by phase-locked fiber lasers in dispersion-managed media \cite{mole,mole2,mole3,mole4,mole5,mole6}.  Molecules can be generated in different ways and one of the significant routes to achieve them is by controlling the parameters of optical/matter-wave solitons \cite{molebec,mole,mole2,mole3,mole4,mole5,mole6, pra2022}. Such questions naturally arise here to understand molecules in bosonic condensates, which are not yet studied for multicomponent BECs in general and for spinor BECs in particular.
	
	Motivated by the above interesting observations and for further understanding, we devote our objective to studying the dynamics of bright matter-wave SMs in spin-1 Bose-Einstein condensates with different forms of modulated nonlinearities. First, we provide certain essential descriptions of the non-autonomous spinor model under consideration in the next section \ref{sec-model} along with the methodology used for the present study. Section \ref{sec-revisit} contains a brief revisit on the soliton dynamics in the underlying autonomous system. A categorical analysis of the generation of spinor matter-wave bright SMs with explicit solutions and their dynamics under modulation by non-autonomous nonlinearities is given in section \ref{sec-mole}. The final section \ref{sec-conclu} is allotted for conclusions along with certain future directions. 
	
	\section{The Mathematical Model and Method}\label{sec-model}
	We consider the following three-component non-autonomous Gross-Pitaevskii (3-NAGP) equations describing the dynamics of an ultra-cold dilute gas of optically trapped spin $F=1$ bosonic condensates \cite{PGK-book,Pethick-book,NatPRL,Ohta,PhysRep}:
	\bes\bea
	&&i\frac{\partial \psi_{+1}}{\partial t} = \left(-\frac{1}{2}\frac{\partial^2}{\partial x^2}+V_{\text{ext}}(x,t)\right)\psi_{+1}+[c_0(t)+c_2(t)](|\psi_{+1}|^2+2|\psi_{0}|^2)\psi_{+1} +2c_2(t)\psi_0^2\psi_{-1}^* \nonumber\\ && \hspace{6.05cm} +[c_0(t)-c_2(t)]|\psi_{-1}|^2\psi_{+1},\\
	&&i\frac{\partial \psi_{0}}{\partial t} =\left(-\frac{1}{2}\frac{\partial^2}{\partial x^2}+V_{\text{ext}}(x,t)\right)\psi_{0}+[c_0(t)+c_2(t)](|\psi_{+1}|^2+|\psi_{-1}|^2)\psi_{0} +2c_0(t)|\psi_0|^2\psi_{0}\nonumber\\ && \hspace{6.05cm} +2c_2(t)\psi_{-1}\psi_{0}^*\psi_{+1},\\
	&&i\frac{\partial \psi_{-1}}{\partial t} = \left(-\frac{1}{2}\frac{\partial^2}{\partial x^2}+V_{\text{ext}}(x,t)\right)\psi_{-1}+[c_0(t)+c_2(t)](|\psi_{-1}|^2+2|\psi_{0}|^2)\psi_{-1} +2c_2(t)\psi_0^2\psi_{+1}^*\nonumber\\ &&\hspace{6.05cm} +[c_0(t)-c_2(t)]|\psi_{+1}|^2\psi_{-1},
	\eea\label{na3gp}\ees
	where $\psi_{\pm 1,0}$ denote wave functions of the three spin components with spin projection $m_F=\pm 1,0$, $t$ and $x$ are time and spatial co-ordinates, and $V_{\text{ext}}(x,t)$ represents an external trapping potential which can be chosen among different forms like harmonic, double-well, optical lattice potential, etc. \cite{serkin,tkjpanonat,prebabu}. Here the mean-field (spin-independent) interaction $c_0(t) = {2(a_0(t) + 2a_2(t))}/{3}$ and the spin-dependent interaction $c_2(t) = {2(a_2(t) - a_0(t))/}{3}$ arise as function of time-varying $s$-wave scattering lengths $a_0(t)$ and $a_2(t)$ that can be tuned by Feshbach resonance mechanism \cite{NatPRL}. The above 3-NAGP equation is generally non-integrable, but its constant-nonlinearity counterparts become integrable for certain special choices of mean-field  and the spin-dependent interactions $c_0 = c_2 = c$ and $c_2=0$ \cite{tkjmp}. Here the former choice includes the contribution of spin-mixing nonlinearities, while it is absent in the latter and reduces to the standard Manakov/coupled GP type equation \cite{tk17}. The nonlinearity coefficient $c$ can be either a positive or negative real constant representing repulsive or attractive spinor condensates for which soliton solutions and their dynamics are reported without the external potential \cite{Ohta,tkjmp}. 
	{ As our focus is to study soliton molecules using the exact solution of the considered non-autonomous spinor BEC model, we choose both mean-field and spin-exchange interaction coupling coefficients $c_0(t)$ and $c_2(t)$ (which depend on the $s$-wave scattering lengths $a_0(t)$ and $a_2(t)$) to be equal, i.e., $c_0(t)=c_2(t)=c(t)$ based on the mentioned integrability choice along with the similarity transformation result in homogeneous spinor system (without potential). Due to this specific consideration, the role of miscible and immiscible behaviour does not play any important role in the present study dealing with the dynamics of solitons with an explicit solution.} 
	The above model has its equivalent counterparts in optical systems with/without coherent four-wave mixing effects in addition to self- and cross-phase modulation nonlinearities \cite{Kiv-book,RK97,tkopt,tkopt2,Park,ksjpa,ksjmp,kspram15}. 
	
	Now, we consider the following constant-coefficient (autonomous) three-coupled Gross-Pitaevskii (3-GP) equations \cite{Ohta,tkjmp,tkwcna,tkpla14}: 
	\bes \bea 
	&&iQ_{1,T}+Q_{1,XX}+2(|Q_1|^2+2|Q_2|^2){Q_1} + 2 Q_2^2 Q_3^*=0,\\
	&&iQ_{2,T}+Q_{2,XX}+2(|Q_1|^2+|Q_2|^2+|Q_3|^2){Q_2} + 2 Q_1 Q_3 Q_2^*=0,\\
	&&iQ_{3,T}+Q_{3,XX}+2(2|Q_2|^2+|Q_3|^2){Q_3} + 2 Q_2^2 Q_1^*=0.
	\eea \label{3agp} \ees 
	where $Q_j,~j=1,2,3,$ are spin components of the autonomous system while $T$ and $X$ are transformed time and space coordinates, respectively. 
	To attain our objective, studying the dynamics of matter-wave SMs in the spinor condensates, we convert the 3-NAGP model (\ref{na3gp}) for an attractive nonlinearity case $c_0(t)=c_2(t)=-c(t)$, where $c(t)$ is a positive function with one-dimensional harmonic potential $V_{\text{ext}}(x,t)=\frac{1}{2}\Omega^2(t)x^2$ to the above autonomous 3-GP equation (\ref{3agp}) by implementing the following similarity transformation:
	\bes\bea
	&&\begin{pmatrix} 
		\psi_{+1}(x,t)\\
		\psi_{0}(x,t)\\
		\psi_{-1}(x,t)
	\end{pmatrix} 
	= \xi_1\sqrt{{c}(t)}
	\begin{pmatrix} 
		Q_1(X,T)\\
		Q_2(X,T)\\
		Q_3(X,T)
	\end{pmatrix} 
	e^{i\theta(x,t)},
	\eea where 
	\bea
	&&\theta(x,t) = -\frac{x^2}{2}\frac{d}{dt}(\ln c(t)) +  2\xi_1^2 \xi_2 \left({c(t)}x - \xi_1^2 \xi_2 \int_0^t {c(t)}^2 dt\right),\\
	&&X(x,t) = ~\sqrt{2}\xi_1 \left({c(t)} x - 2\xi_1^2 \xi_2 \int_0^t {c(t)}^2 dt\right),\\
	&&T(t) = \xi_1^2 \int_0^t {c(t)}^2 dt,
	\eea
	along with the condition in the form of a Riccati-type equation
	\bea
	\Omega^2(t)=-\frac{1}{2c(t)}\frac{d^2c(t)}{dt^2}+\frac{1}{c(t)^2}\left(\frac{dc(t)}{dt}\right)^2.\label{ricatti}
	\eea 
	\label{str}\ees
	The above-mentioned solution (\ref{str}) is a phase-locked solution as all three components possess the same phase $\theta$, which has a quadratic dependence on $x$. Note that for constant nonlinearity coefficient $c$, the phase $\theta$ has a linear dependence on $x$ and $t$. 
	The constraint condition (\ref{ricatti}) is nothing but the Riccati equation and it determines the relationship between the external potential $V_{ext}(x,t)$ and nonlinearity $c(t)$. The above transformation contains two arbitrary (similarity) parameters $\xi_1$ and $\xi_2$ that will play an important role in the dynamics of the associated nonlinear wave. Through the above transformations, one can analyze the dynamics of any type of nonlinear waves, localized or travelling, associated with the non-autonomous model (\ref{na3gp}), once we know the solutions of its autonomous counterpart (\ref{3agp}). Thus, in the following sections, we can study the impact of non-autonomous nonlinearities in the formation and evolution of bound SMs of the 3-NAGP system (\ref{na3gp}) by using the transformation (\ref{str}). 
	
	{ Here it is important to mention that the above similarity transformation (\ref{str}) has certain limitations as it cannot be applied to all types of nonlinear evolution equations (PDEs) due to the occurrence of a number of constraints which hinder the process of getting more general solutions. In the present work, we have the Riccati type equation (\ref{ricatti}) as the restriction during the solution process, which determines the form of corresponding potential based on the chosen nonlinearity coefficient. Though this Riccati equation is more physical, we can still say that it is an unavoidable limitation in the present similarity transformation. Further, the constructed similarity transformation  (\ref{str}) cannot be applied straightforwardly (as it is) to other nonlinear equations (especially the present spin-1 model (1)) with both spatio-temporal-dependent nonlinearities (i.e., $c(x,t)$) and pure spatial-dependent nonlinearities ($c(x)$). These problems require further tailoring in the transformation to every model equation, which can help us to obtain explicit solutions of our interest. } 
	
	{ Beyond the above limitations, the present similarity transformation has certain advantages that add significant novelty to the work, as listed below. 
		(i) The generalized similarity transformation to the three-component spin-1 BEC system by incorporating physically interesting nonlinearity modulations is new compared with similar approaches in other (scalar and two-coupled) models.
		(ii) A more general nonlinearity variation is considered in the form of the Jacobi elliptic function, which encompasses the trignometric and hyperbolic type nonlinearity variations. Such modulation dynamics can be physically realized in an optical lattice potential trap. 
		(iii) Success in similarity transformation is the achievement of required solutions with less number of constraints. Fortunately, in the present work, we have achieved with single constraint only, which also becomes the physically significant Riccati equation. }
	
	\section{Bright Matter-Wave Solitons: Revisit}\label{sec-revisit} 
	As mentioned in the introduction, we can understand that the exact soliton solutions of both 3-NAGP and 3-GP models have been obtained by using different methods along with a detailed discussion on their propagation and interaction dynamics \cite{Ohta,tkpla14,tkwcna}. Though we intend to deal with SMs alone, it is essential to emphasize certain entities of fundamental solitons associated with the considered spinor BEC system. To facilitate the understanding of the formation of SM and for completeness, we briefly revisit the nature of solitons arising in the integrable autonomous 3-GP equations using the previous results. 
	By adopting the non-standard Hirota's bilinearization method \cite{tkwcna,tkpla14,Hirota-book,gilson}, we can write the matter-wave bright one-soliton solution for the autonomous 3-GP equation (\ref{3agp}) in its explicit form as below.
	\bea
	&&Q_{j}(X,T) = \frac{\alpha_1^{(j)} e^{\eta_1}+e^{2\eta_1+\eta_1^*+\delta_{11}^{(j)}}}{1+e^{\eta_1+\eta_1^*+R_1}+e^{2\eta_1+2\eta_1^*+\epsilon_{11}}},\quad j=1,2,3, \label{one-sol}
	\eea
	where $\eta_1 = k_1(X+ik_1T)$, $e^{R_1}=\frac{{(|\alpha_1^{(1)}|^2+2|\alpha_1^{(2)}|^2+|\alpha_1^{(3)}|^2)}}{(k_1+k_1^*)^2}$, $e^{\epsilon_{11}}=\frac{|\Gamma_1|^2}{(k_1+k_1^*)^4}$, $e^{\delta_{11}^{(j)}}=\frac{(-1)^{j+1} \alpha_1^{(4-j)*} \Gamma_1}{(k_1+k_1^*)^2}$, with $j=1,2,3$, and $\Gamma_1=\alpha_1^{(1)}\alpha_1^{(3)}-(\alpha_1^{(2)})^2$. Here $k_1$, $\alpha_1^{(1)}$, $\alpha_1^{(2)}$, $\alpha_1^{(3)}$ are arbitrary complex parameters, which characterize the dynamics of bright one-soliton.  
	\begin{figure}[hb]
		\centering\includegraphics[width=0.477\linewidth]{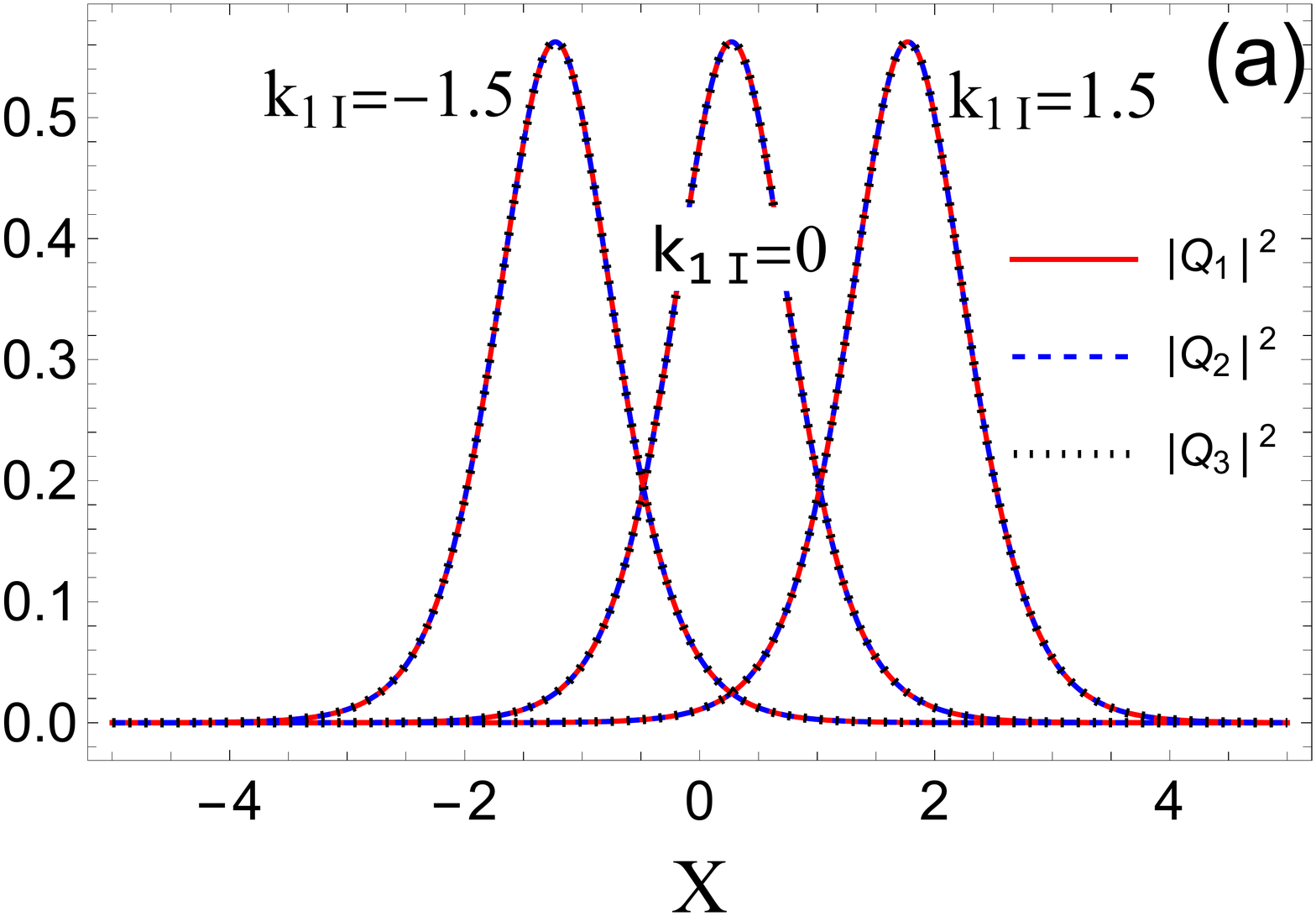}\qquad \includegraphics[width=0.468\linewidth]{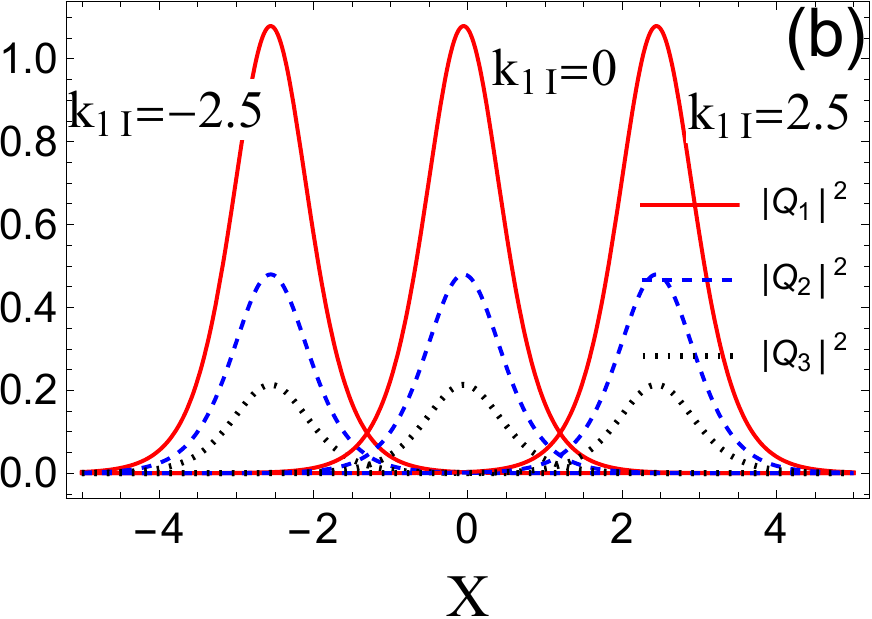}
		\caption{Perfect/Symmetric bell (sech) type ferromagnetic matter-wave bright soliton resulting when $\Gamma_1=\alpha_1^{(1)}\alpha_1^{(3)}-(\alpha_1^{(1)})^2=0$ for three different velocities with $k_{1R}=1.5$ at a particular time $T=0.2$. (a) Degenerate FSs (equal density $|Q_1|=|Q_2|=|Q_3|$) for $\alpha_1^{(1)}=\alpha_1^{(2)}=\alpha_1^{(3)}=1.0$, and (b) Non-degenerate FSs (distinct densities $|Q_1|\neq |Q_2|\neq |Q_3|$) for $\alpha_1^{(1)}=2.25,\alpha_1^{(2)}=1.5,$ \& $\alpha_1^{(3)}=1.0$.}
		\label{fs1sol}
	\end{figure}
	
	\begin{figure}[ht]
		\centering\includegraphics[width=0.4863\linewidth]{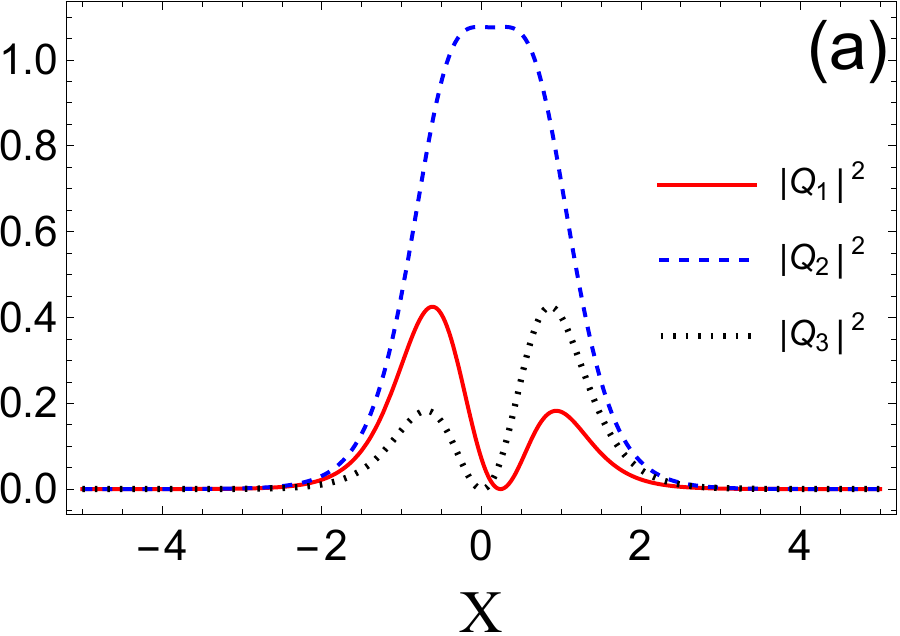}\qquad \includegraphics[width=0.4863\linewidth]{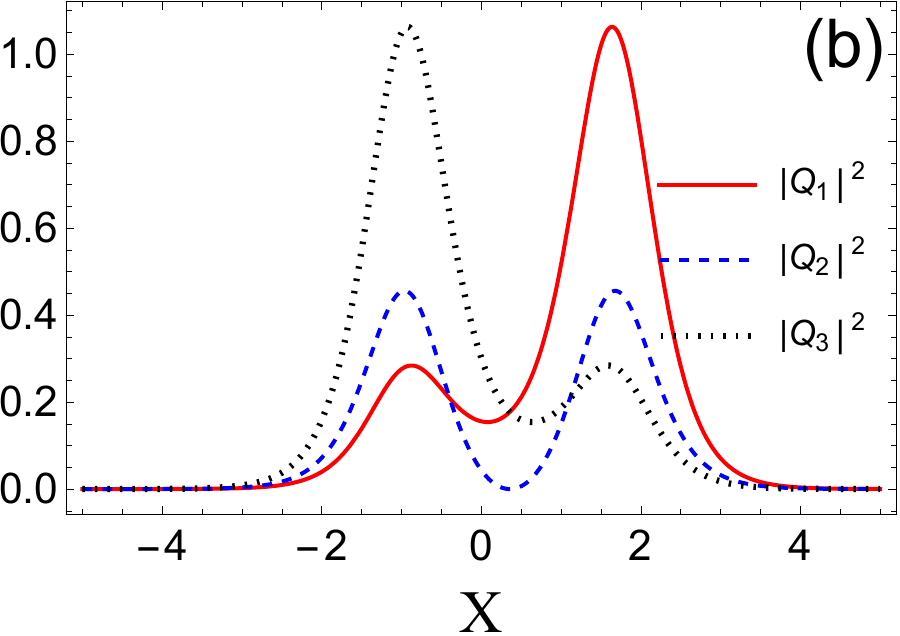} \includegraphics[width=0.4863\linewidth]{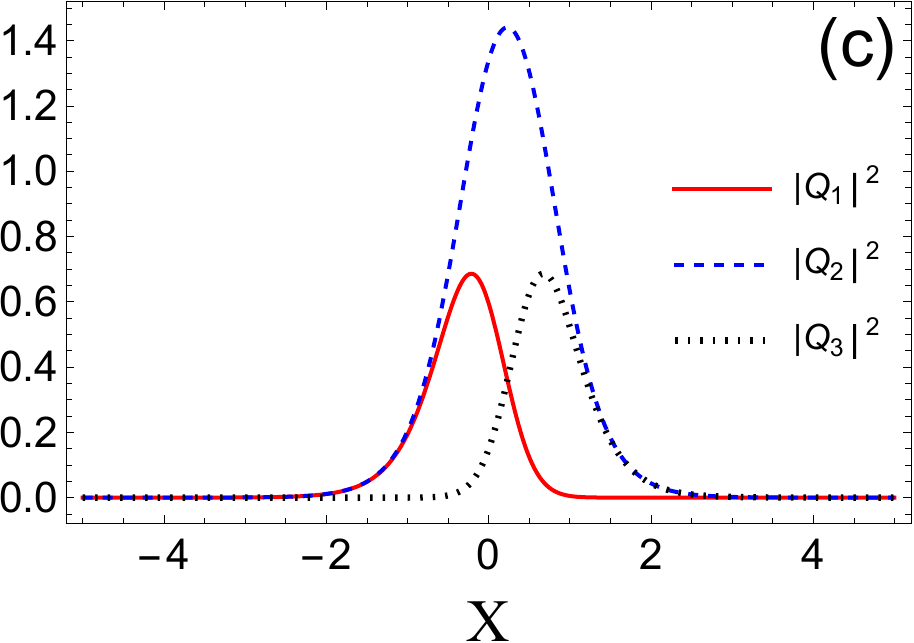}
		\caption{Polar solitons (PSs) with different types of symmetric/asymmetric profiles resulting when $\Gamma_1=\alpha_1^{(1)}\alpha_1^{(3)}-(\alpha_1^{(1)})^2 \neq 0$ with $k_1=1.5-0.5i$ at a particular time $T=0.2$. (a) Asymmetric double-hump (in $Q_1$ \& $Q_3$) and symmetric flat-top (in $Q_2$) PSs for $\alpha_1^{(1)}=\sqrt{2},\alpha_1^{(2)}=1.67,$ \& $\alpha_1^{(3)}=1.0$). (b) Asymmetric double-hump (in $Q_1$ \& $Q_3$) and symmetric double-hump (in $Q_2$) PSs for $\alpha_1^{(1)}=1,\alpha_1^{(2)}=1.35,$ \& $\alpha_1^{(3)}=2.0$. (c) Asymmetric shifted single-hump (in $Q_1$ \& $Q_3$) and symmetric single-hump (in $Q_2$) PSs for $\alpha_1^{(1)}=\alpha_1^{(2)}=1.0,$ \& $\alpha_1^{(3)}=0.0$.}
		\label{ps1sol}
	\end{figure}
	
	Solitons derived from the above solution can be classified into ferromagnetic soliton (FS: non-zero total spin) and polar soliton (PS: zero total spin). These PS and FS arise for the choice $S\neq 0$ and $S=0$, respectively, where the auxiliary function $S=\Gamma_1 e^{2\eta_1}$ \cite{Ohta,tkwcna,tkpla14}. Specifically, the contribution resulting from the spin-mixing nonlinearity vanishes ($\Gamma_1=0$) in the case of FSs and results in a certain non-zero value of total spin. Its explicit solution becomes a standard vector soliton with a symmetric $sech$ (bell) type profile with the same/equal or distinct/non-equal amplitude among the three components. On the other hand, the presence of spin-mixing nonlinearity makes the total spin vanish and generates PSs admitting complex solution form and various types of profiles starting from symmetric single-hump, double-hump, flat-top to asymmetric single/double-hump structures for different choices of $\alpha_1^{(j)}$ parameters. To elucidate these features, we have shown certain soliton behaviour for bright FSs and PSs in Fig. \ref{fs1sol} and Fig. \ref{ps1sol}. For a more detailed mathematical and physical interpretations, one can refer to \cite{Ohta,tkpla14}. Here, we have to note that apart from the shape/nature of profiles, which is defined by $\alpha_1^{(j)}$ alone, the remaining characteristics of solitons such as amplitude, velocity, and central position are determined by the parameters $k_1=k_{1R}+i k_{1I}$ in addition to $\alpha_1^{(j)}$, $j=1,2,3$. To be precise, $k_{1R}$ and $\alpha_1^{(j)}$ combined to influence the amplitude of the soliton in each component, while $k_{1I}$ controls the velocity in all components uniformly. Here and in the following, the subscripts $R$ and $I$ represent the real and imaginary parts of the corresponding parameter, respectively. Importantly, the direction of soliton propagation becomes right-moving when $k_{1I}>0$ (left-moving when $k_{1I}<0$) with appropriate speed given by the magnitude of $|k_{1I}|$. This velocity parameter will play a pivotal role in the formation of bound SMs, which is the main objective of the manuscript, and its detailed discussion is given in the next section.   
	
	\section{Bright Matter-Wave Soliton Molecules}\label{sec-mole} 
	In order to understand the dynamics of bound SMs of any nonlinear dynamical system analytically, its exact multi-soliton solutions or at least two soliton solution is required. This section discusses the fundamental SMs comprising two bright matter-wave solitons and explores their dynamics for both autonomous and non-autonomous spinor systems. For this purpose, we consider the matter-wave bright two-soliton solution obtained by the present authors \cite{tkwcna,tkpla14} for 3-NAGP model \eqref{na3gp} and using the above-discussed similarity transformation \eqref{str} in its explicit form as below.
	\bes \bea
	\big(\psi_{+1}(x,t),~\psi_{0}(x,t),~\psi_{-1}(x,t)\big)^T = \xi_1\sqrt{{c}(t)}\big(Q_1(X,T),~Q_2(X,T),~Q_3(X,T)\big)^T e^{i\theta(x,t)},~ j=1,2,3, \label{nonat2sol} \eea  
	where \vspace{-0.63cm} \bea Q_j(X,T) = \frac{G^{(j)}(X,T)}{F(X,T)}, \qquad j=1,2,3. \hspace{3.0cm} \label{2sol} \eea
	\ees
	In the above two-soliton solution, the superscript $T$ denotes the transpose, ${G^{(j)}(X,T)},~j=1,2,3,$ and ${F(X,T)}$ are complex and real functions, respectively, and evolve with respect to the envelopes $\eta_1(X,T)=k_1(X+i k_1 T)$ and $\eta_2(X,T)=k_2(X+i k_2 T)$, representing two matter-wave solitons (say, $S_1$ and $S_2$) and their explicit form is given in \ref{appendix}. We can note that in addition to the terms with $\Gamma_1=\alpha_1^{(1)}\alpha_1^{(3)}-(\alpha_1^{(2)})^2$ and $\Gamma_2=\alpha_2^{(1)}\alpha_2^{(3)}-(\alpha_2^{(2)})^2$ on the parameters corresponding to $S_1$ and $S_2$, there exists another parameter  $\Gamma_3=\alpha_1^{(1)}\alpha_2^{(3)}+\alpha_2^{(1)}\alpha_1^{(3)}-2\alpha_1^{(2)}\alpha_2^{(2)}$ resulting in a combined contribution of both solitons due to spin-mixing nonlinearity. As discussed for the bright one-soliton case in the previous section, the present two-soliton can also be classified into FS and PS with respect to the total density using spin-mixing nonlinearity (auxiliary function) contributions \cite{Ohta,tkpla14}. Note that the FSs admit symmetric single-hump profiles alone, while the PSs can have both symmetric and asymmetric type single-hump/double-hump/flat-top structures for a proper set of parameters. Further, the interaction among two solitons falls under different categories, such as (i) FS with FS, (ii) PS with PS, and (iii) FS with PS, thereby resulting in various interaction behaviours that include (i) elastic and spin-precession interactions of single-hump solitons, (ii) elastic interactions of symmetric double-hump/flat-top and asymmetric single-hump solitons, and (iii) spin-switching interaction of single-hump/double-hump/flat-top PS leaving the symmetric single-hump soliton unaltered. We refrain from providing these interaction dynamics here and one can refer to \cite{tkpla14} for a detailed discussion on the solitons in both autonomous and non-autonomous 3-GP systems (\ref{3agp}) and (\ref{na3gp}).
	
	\begin{figure}[ht]
		\centering\includegraphics[width=0.78\linewidth]{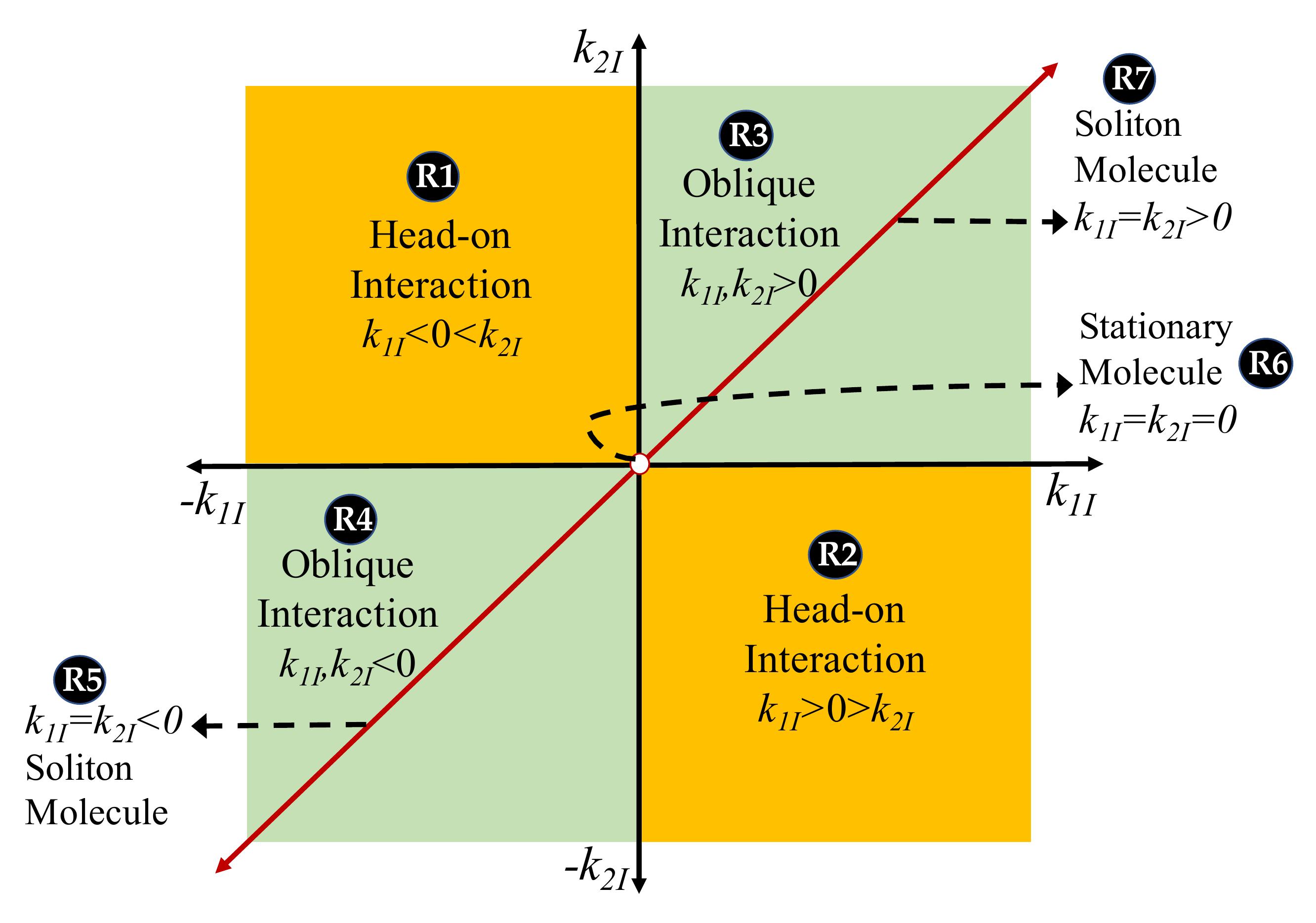} 
		\caption{Schematic representation on the role of velocity-defining parameters $k_{1I}$ and $k_{2I}$ in the interaction and molecule formation among two bright matter-wave solitons. The two parametric regions \textbf{R1} (with $k_{1I}>0>k_{2I}$) and \textbf{R2} (with $k_{1I}<0<k_{2I}$) support head-on interactions, while the regions \textbf{R3} (with $k_{1I},k_{2I}>0$) and \textbf{R4} (with $k_{1I},k_{2I}<0$) exhibit oblique type interactions of bright two-solitons revealing elastic, energy-sharing, and energy-switching characteristics based on the polarization parameters $\alpha_u^{(j)}$. The parametric choice denoted by the red line corresponds to the requirement $k_{1I}=k_{2I}$ for soliton molecule generating interesting breathers (soliton-chain) can be divided into three regions. Regions \textbf{R5} ($k_{1I}=k_{2I}<0$) and \textbf{R7} ($k_{1I}=k_{2I}>0$) represent traveling molecules, while \textbf{R6} ($k_{1I}=k_{2I}=0$) denotes stationary SM. Explicit graphical representations of interactions and molecules are given in Fig. \ref{schematic-2}.}
		\label{schematic}
	\end{figure}
\begin{figure}[ht]
		\centering\includegraphics[width=0.998\linewidth]{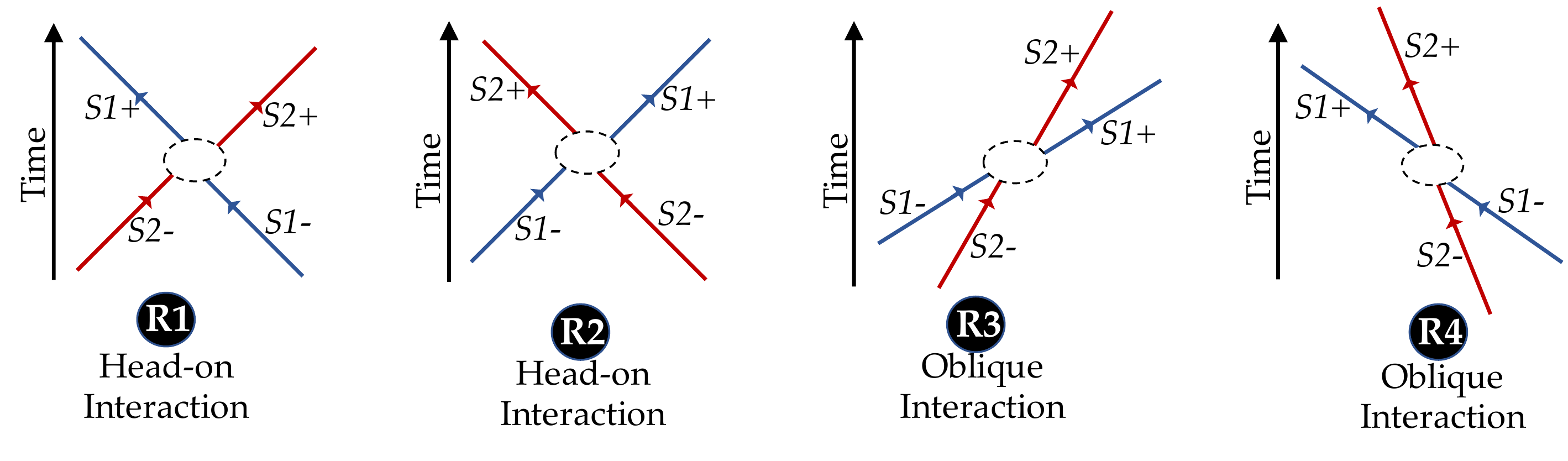}\\ 
		\centering\includegraphics[width=0.85\linewidth]{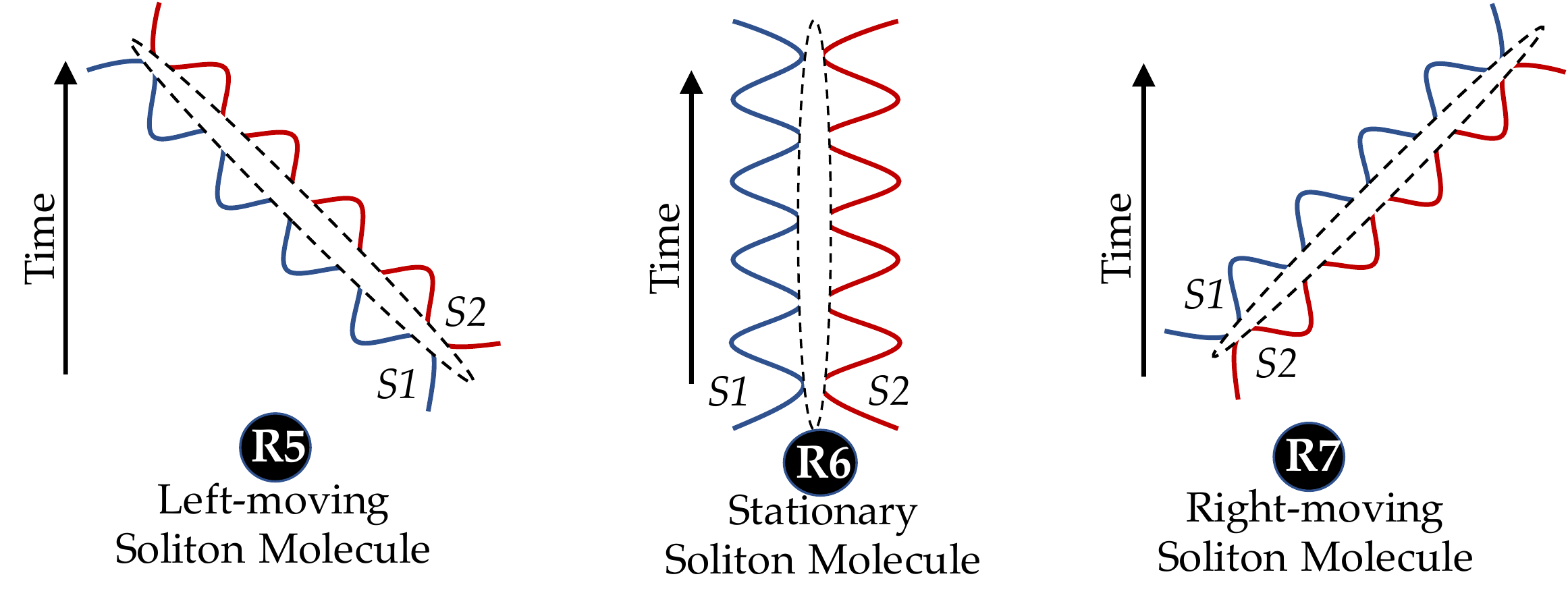}
		\caption{Schematic representation of different possible interactions and molecule formation by two bright matter-wave solitons for various choices of parametric regions given in Fig. \ref{schematic} based on $k_{1I}$ and $k_{2I}$. Here blue and red lines represent soliton 1 and soliton 2, respectively, while the area marked by dashed-black-ellipsoids corresponds to the interaction regime, which is very short for standard soliton interactions, whereas it becomes a very-long regime for soliton molecules. }
		\label{schematic-2}
	\end{figure}

	Here, our focus is to analyze every possibility to generate bright SM and unravel various dynamical features resulting from effective manipulation of the underlying solitons in both autonomous and non-autonomous settings. From the one-soliton solution, we notice the importance of $k_u=k_{uR}+ik_{uI}$ ($u=1,2$), where the real part of the parameter ($k_{uR}$) takes part in defining the amplitude of the profile and the imaginary part ($k_{uI}$) controls the velocity of the constituent soliton. Note that this $k_{uI}$ parameter not only influences the velocity and nature of the interaction (whether it is a head-on or oblique/overtaking interaction) but also accounts for the formation of the molecule. To be specific, the possibility of identifying SMs mainly depends upon the velocity-matching or velocity-resonance property of constituent solitons with a minimum of two fundamental solitons. However, there can exist any number of solitons which shall produce multi-soliton molecules. This represents that when solitons are propagating in a media with the same velocity, they start moving together with mutual and continuous interactions leading to periodic attraction and repulsion lasting throughout the propagation distance. Such bound propagation of solitons is referred to as coalescence of wavenumbers due to the constraint on the parameter $k_u$. Further, one can also generate both stationary SMs and travelling SMs by choosing the associated velocity parameters to be zero and non-zero (either positive or negative constant), respectively, denoted as $k_{1I}=k_{2I}=k_{3I}=\dots=k_{nI}=0$ and $k_{1I}=k_{2I}=k_{3I}=\dots=k_{nI}\neq 0$ with arbitrary $n$ number of solitons. For easy understanding, a schematic representation of the role of $k_{1I}$ and $k_{2I}$ parameters in the interactions and molecule formation involving two bright solitons are shown in Figs. \ref{schematic} and \ref{schematic-2}.
	
	Additionally, these bright bound SMs can be classified into two types based on the central positions of participating solitons, namely molecules with (i) the same and (ii) distinct central positions defined by the $k_{jR}$ parameters. The former results in coinciding or overlapped solitons for $k_{1R}=k_{2R}=k_{3R}=\dots=k_{nR} \neq 0$, where all the available solitons merge together and form a single-soliton-like structure with superior (higher) amplitude and stable profile. However, the latter admits non-coinciding or displaced solitons for the choice $k_{1R} \neq k_{2R}\neq k_{3R}\neq \dots \neq k_{nR}\neq 0$, which leads to the bounded propagation of these multiple solitons with different central positions and they exhibits periodic oscillations in their amplitudes.
	
	In the present spinor 3GP system, we formulate two-soliton molecule through a velocity resonance $k_{1I}=k_{2I}$ with same ($k_{1R} = k_{2R}$) or different ($k_{1R} \neq k_{2R}$) central positions. Particularly, we deal with molecules formed under different pairs of bright solitons, namely (i) FS$\times$FS $\Rightarrow$ Ferromagnetic Soliton Molecule (FSM), (ii) PS$\times$PS $\Rightarrow$ Polar Soliton Molecule (PSM), and (iii) FS$\times$PS $\Rightarrow$ mixed Ferromagnetic-Polar Soliton Molecule (FPSM) by appropriately tailoring the arbitrary parameters $\alpha_u^{(j)}$ ($u=1,2$ and $j=1,2,3$). Interestingly, we can generate both stationary and moving SMs with coinciding and non-coinciding central positions in the autonomous (\ref{3agp}) and non-autonomous (\ref{na3gp}) systems showing different characteristics due to diverse profile structures. For bringing out their salient features, we study these molecules and their dynamics with constant and temporally-varying nonlinearities for each of these cases in detail. 
	
	\subsection{\bf Ferromagnetic Soliton Molecule (FSM)}
	We can obtain ferromagnetic type two-soliton molecules when the spin-mixing nonlinearity vanishes. In Refs. \cite{tkpla14,tkwcna}, it is shown that the FSs admit elastic and spin-precession interactions which results for the choice $\Gamma_3=0$ and $\Gamma_3 \neq 0$, respectively along with $\Gamma_1=\Gamma_2=0$. Here also, we can form molecules under these two situations as we discussed below. To understand the facts clearly, we write the explicit expression for $G^{(j)}$ and $F$ representing FSM with two bright solitons, resulting for the choice $\Gamma_1=\Gamma_2=0$ and $k_{2I}=k_{1I}$ by keeping central positions distinct with $k_{1R}\neq k_{2R}$, due to which several terms vanish and takes a simple form as given below.
	\bes\bea 
	\hspace{-1.5cm}G^{(j)}_{FSM}&=&\alpha _1^{(j)} e^{\eta _1} +\alpha _2^{(j)} e^{\eta _2} +e^{\eta_1+\eta_2+\eta_1^*+\delta_1^{(j)}}+e^{\eta_1+\eta_2+\eta_2^*+\delta_2^{(j)}},\quad j=1,2,3, \label{Gfsm}\\
	\hspace{-1.5cm}F_{FSM}&=&1+e^{2\eta_{1R}+R_1}+e^{2\eta_{2R}+R_2}+e^{\eta_1+\eta_2^*+\delta_0}+e^{\eta_2+\eta_1^*+\delta_0^*}+e^{2(\eta_{1R}+\eta_{2R})+{R_3}}.
	\eea 
	The above form can be rewritten in terms of trigonometric-hyperbolic functions, from which the explicit autonomous and non-autonomous FSM solutions take the form as given below.
	\bea 
	Q_j(X,T)=\frac{e^{\frac{\delta_2^{(j)}+l_1^{(j)}}{2}}\cosh\left({\eta_{2R}}+N_{2R}^{(j)}+iN_{2I}^{(j)}\right)e^{i\eta_{1I}} + e^{\frac{\delta_1^{(j)}+l_2^{(j)}}{2}}\cosh\left({\eta_{1R}}+N_{1R}^{(j)}+iN_{1I}^{(j)}\right)e^{i\eta_{2I}}}{e^{\frac{R_1+R_2}{2}}\cosh\left({\eta_{1R}-\eta_{2R}+\frac{R_1-R_2}{2}}\right) + e^{\frac{R_3}{2}}\cosh\left({\eta_{1R}+\eta_{2R}+\frac{R_3}{2}}\right) +  e^{\delta_{0R}} \cos{(\eta_{1I}-\eta_{2I}+\delta_{0I})}},\label{afsm}\\
	\left(\psi_{+1}(x,t),~ \psi_{0}(x,t),~\psi_{-1}(x,t)\right)^T = \xi_1\sqrt{{c}(t)}~ (Q_1,~Q_2,~Q_3)^T e^{i\theta(x,t)},\qquad j=1,2,3.\qquad\qquad\label{nafsm}
	\eea \ees 
	Here $N_{1R}^{(j)}=\frac{\delta_{1R}^{(j)}-l_{2R}^{(j)}}{2}$, $N_{1I}^{(j)}=\frac{\delta_{1I}^{(j)}-l_{2I}^{(j)}}{2}$, $N_{2R}^{(j)}=\frac{\delta_{2R}^{(j)}-l_{1R}^{(j)}}{2}$, $N_{2I}^{(j)}=\frac{\delta_{2I}^{(j)}-l_{1I}^{(j)}}{2}$,  $\eta_{uR}=k_{uR}(X-2k_{1I} T)$ and $\eta_{uI}=k_{1I} X+(k_{uR}^2 - k_{1I}^2)T$, $u=1,2$. Note that though $k_{1I} - k_{2I} = 0$, $\eta_{1I} - \eta_{2I} \neq 0$ and $\eta_{1R} - \eta_{2R} \neq 0$. 
	Also, $e^{l_u^{(j)}}=\alpha_u^{(j)}$ and ${\delta_u^{(j)}},~u=1,2$, are complex functions. 
	
	The autonomous FSM is now characterized by four (out of six) spin-polarization parameters $\alpha_u^{(j)}$ and three arbitrary real constants $k_{1R},~k_{2R}$ and $k_{1I}$. By analyzing the above form, we can observe that the localized FSM admits oscillation in its amplitude along with periodic attraction and repulsion in localization due to the simultaneous existence of hyperbolic and trigonometric functions. It is essential to note the significance of bound molecules compared to that interacting solitons. During an elastic interaction, one can obtain single-maximum/peak amplitude at the one-time interaction regime, after which both solitons pass through each other and regain their initial identities except for a phase shift. However, in molecules, both solitons interact periodically without passing through each other and it continues indefinitely by generating a peak/maxima at every interaction. This is driven by the third periodic term appearing in the denominator of Eq. (\ref{afsm}), which becomes a pure cosine function along $T$ alone with periodicity proportional to $k_{1R}^2-k_{2R}^2$. Thus, based on the magnitude $k_{1R}^2-k_{2R}^2$, which represents the difference in the central positions between the two participating solitons, we can control the period of oscillations and induce asymmetry between the two oscillating profiles. Suppose if we choose the same central position by taking $k_{2R}=k_{1R}$, we can eliminate these oscillations as the term with periodic functions vanishes completely, as the present choice makes $\eta_{2}=\eta_{1}$ and hence the corresponding simplified solution structure can be reduced from (\ref{Gfsm})-(\ref{afsm}) as given below. 
	\bea 
	Q_j(X,T)=\frac{e^{\Delta_1^{(j)}} e^{\eta _1} +e^{\Delta_2^{(j)}}e^{2\eta_1+\eta_1^*}}{1+e^{\Delta_3}e^{2\eta_{1R}}+e^{4\eta_{1R}+{R_3}}}\equiv \frac{e^{\frac{\Delta_1^{(j)}+\Delta_2^{(j)}}{2}}\cosh({\eta_{1R}}+\frac{\Delta_2^{(j)}-\Delta_1^{(j)}}{2})e^{i\eta_{1I}}}{e^{\frac{R_3}{2}}\cosh\left({2\eta_{1R}+\frac{R_3}{2}}\right) + \frac{1}{2}e^{\Delta_3}}, \qquad j=1,2,3,
	\eea 
	where $e^{\Delta_1^{(j)}}=\alpha_1^{(j)}+\alpha_2^{(j)}$, $e^{\Delta_2^{(j)}}=e^{\delta_1^{(j)}}+e^{\delta_2^{(j)}}$, and $e^{\Delta_3}=e^{R_1}+e^{R_2}+e^{\delta_0}+e^{\delta_0^*}$. This can be related to the one-soliton solution (\ref{one-sol}) and reveals the expected merger of two solitons into a stable propagation of a one-soliton-like structure. However, when the central positions are different, the above form will be invalid and turns more complex version (\ref{afsm}) as $k_{1R}\neq k_{2R}$ and $\eta_{2}\neq \eta_{1}$. Thus the resulting solution admits different profiles, from single-hump to double-hump profiles across all three components. Additionally, the SMs exhibit breathing along the amplitude and propagation with the periodicity of oscillations proportional to $k_{1R}^2-k_{2R}^2$. 
	
\begin{figure}[h]
		\centering {~~\qquad (a) \hfill (b) \hfill\quad~~}\\
		\centering\includegraphics[width=0.405\linewidth]{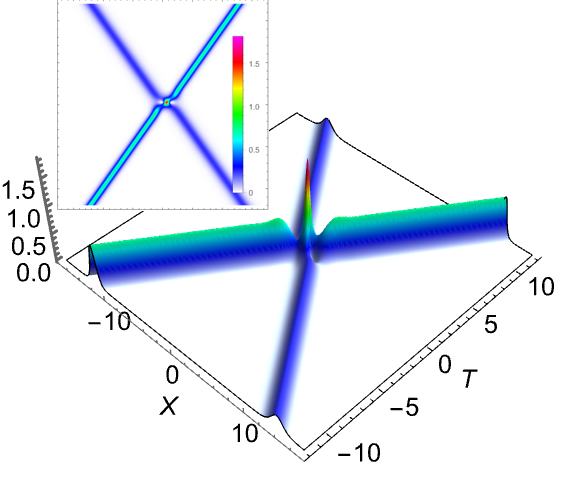}\quad \includegraphics[width=0.405\linewidth]{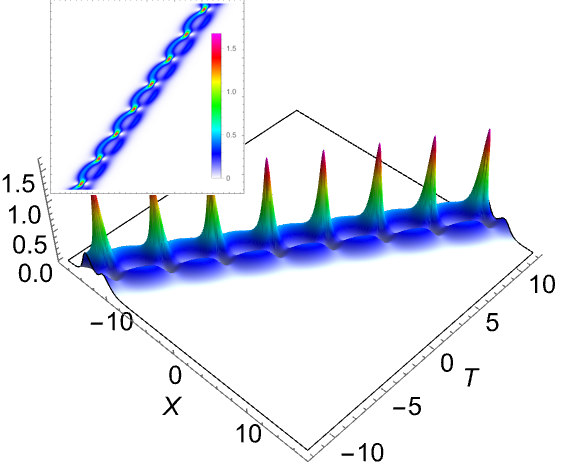}\\
		\centering {~~\qquad (c) \hfill (d) \hfill\quad~~}\\
		\includegraphics[width=0.405\linewidth]{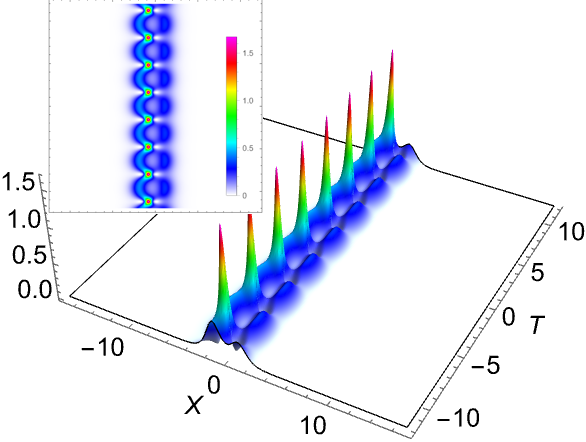}\quad \includegraphics[width=0.405\linewidth]{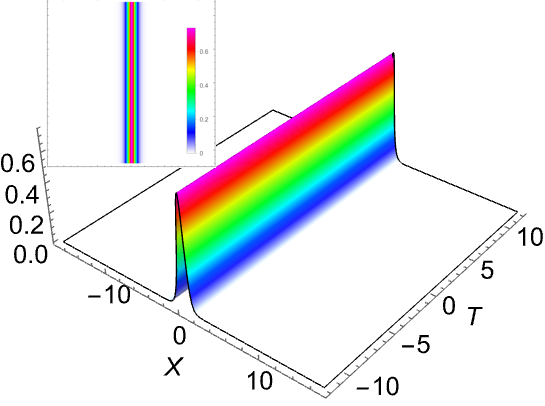}
		\caption{Degenerate type symmetric FSs ($|Q_1|=|Q_2|=|Q_3|$) undergoing standard elastic interaction (top-left (a): $k_1=1+0.5i$ \& $k_2=1.7-0.5i$) and their transformation by velocity resonance leading to the generation of moving FSM (top right (b): $k_1=1+0.5i$ \& $k_2=1.7+0.5i$), stationary FSM (bottom-left (c): $k_1=1+0.0i$ \& $k_2=1.7+0.0i$) and a merged single-soliton-like FSM with stable profile (bottom-right (d): $k_1=k_2=1.7+0.0i$) for $\alpha_1^{(1)}=\alpha_1^{(2)}=\alpha_1^{(3)}=0.2$ and $\alpha_2^{(1)}=\alpha_2^{(2)}=\alpha_2^{(3)}=0.3$.}
		\label{fig-fsm-sym1}
	\end{figure}

	In order to highlight the above discussion, we have graphically depicted the evolution of degenerate and symmetric FSs in Fig. \ref{fig-fsm-sym1}. First, we have shown their standard elastic type head-on interaction due to opposite velocities ($k_{1I}=-k_{2I}=0.5$) in Fig. \ref{fig-fsm-sym1}(a) and how they can be transformed to a moving/travelling FSM for equal-velocities (velocity resonance)$k_{1I}=k_{2I}=0.5$ in Fig. \ref{fig-fsm-sym1}(b). Also, we have shown the stationary FSM having zero-velocity $k_{1I}=k_{2I}=0.0$ with non-coinciding central positions resulting in breathing propagation in Fig. \ref{fig-fsm-sym1}(c). Finally, in Fig. \ref{fig-fsm-sym1}(d), the soliton molecule with coinciding centers showcasing the stable single-soliton-like pattern is given. We can observe that the amplitude of the periodically appearing excited peaks is much higher than that of individual/merged solitons.
	
\begin{figure}[h]
		\centering\includegraphics[width=0.43\linewidth]{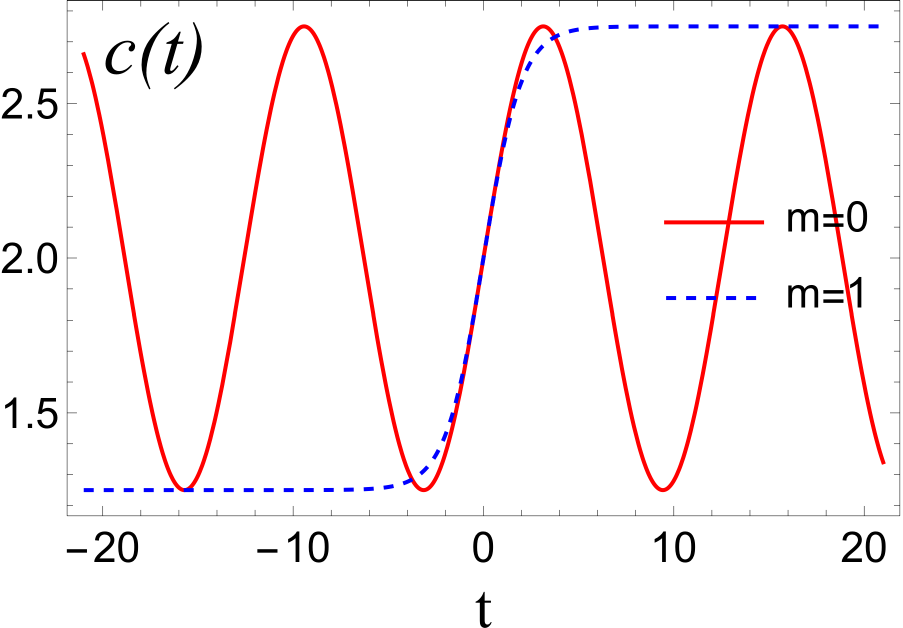}\qquad \includegraphics[width=0.48433\linewidth]{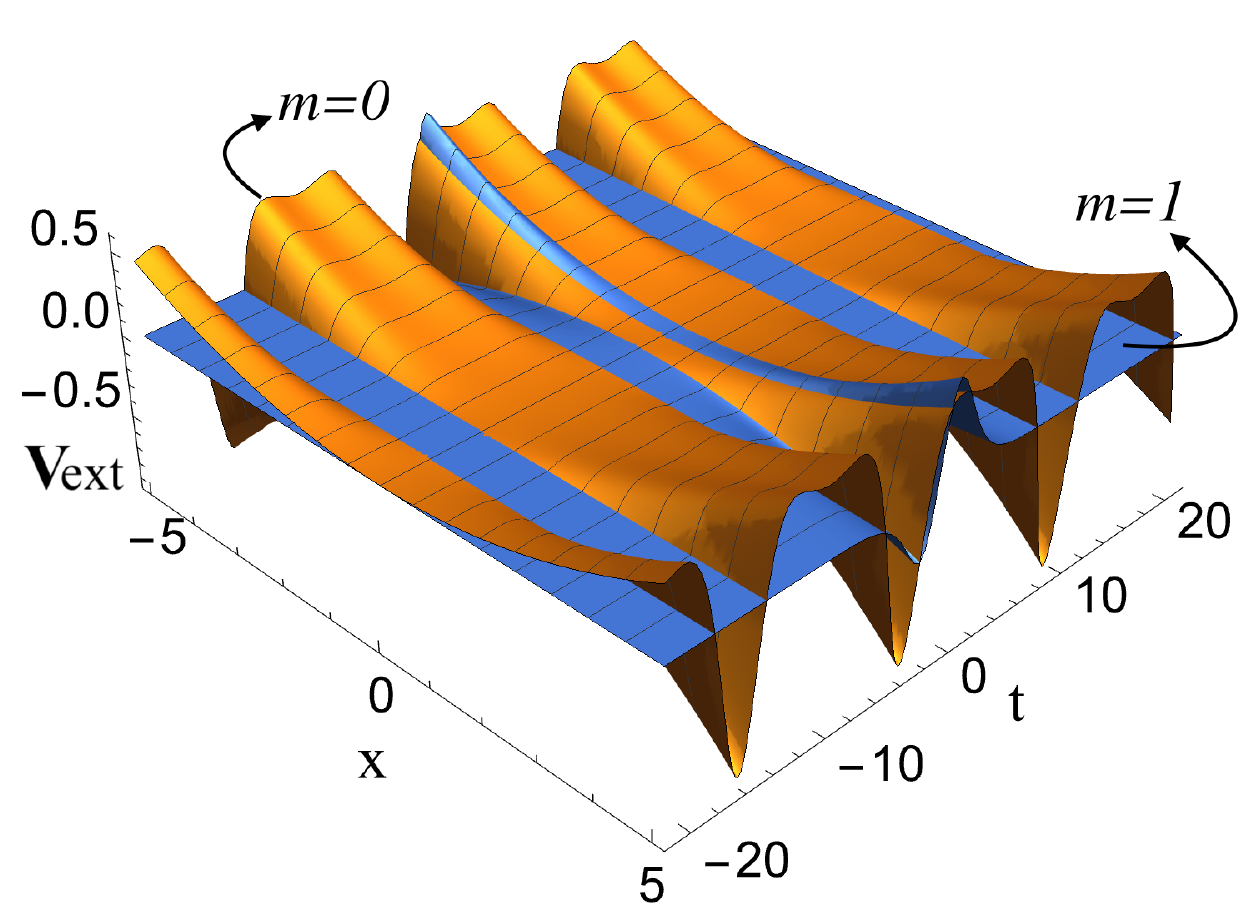}
		\caption{Nature of two forms of nonlinearities  $c(t)$ (left) representing a periodic ($m=0$) and kink-like ($m=1$) behaviour for $p_1=2.0$, $p_2=0.75$, $p_3=0.5$ and $p_4=0.0$ in Eq. (\ref{nonlin}). The associated form of trapping potentials $V_{ext}(x,t)$ are given in the right panel.} 
		\label{fig-pot}
	\end{figure}

	On the other hand, the non-autonomous FSMs are governed by two real constants $\xi_1$ and $\xi_2$ along with the crucial arbitrary nonlinearity function $c(t)$ in addition to the arbitrary parameters $\alpha_u^{(j)}$, $k_{1R},~k_{2R}$ and $k_{1I}$. Mainly, the localized structure of FSM experiences modulations driven by the nonlinearity $c(t)$ and can be chosen among diverse possibilities. Every form of adopted $c(t)$ decides the external potential suitably through the condition (\ref{ricatti}). Additionally, change in the velocity/orientation of molecule is described by the combined contribution from $c(t)$, $\xi_1$ and $\xi_2$, while its amplitude is decided by $c(t)$ and $\xi_1$ alone that we can understand from the similarity relations given in section \ref{str}. To demonstrate the above features, we consider the following simple Jacobi elliptic function for non-autonomous nonlinearity:
	\bea
	c(t)=p_1+p_2~ \mbox{sn}(p_3 t+p_4, m),\quad 0 \leq m \leq 1, \label{nonlin}
	\eea
	where $p_j,~j=1,2,3,4,$ are arbitrary real constants and $m$ is the elliptic modulus parameter. The above form of $c(t)$ encompasses two types of modulated nonlinearities namely periodic (sine) and kink-like (tanh) effects with respect to $t$ for two extreme values of elliptic modulus parameter, $m=0$ and $m=1$, respectively. Also, we can choose constant nonlinearity with magnitude $p_1$ for $p_2=0$. We have shown the nature of the above nonlinearity and the associated potential in Fig. \ref{fig-pot}. Apart from the above simple form of nonlinearity (\ref{nonlin}), one can proceed with several other forms and their superposed forms with appropriately defined external potentials. 
	
	\begin{figure}[h]
		\centering\includegraphics[width=0.34\linewidth]{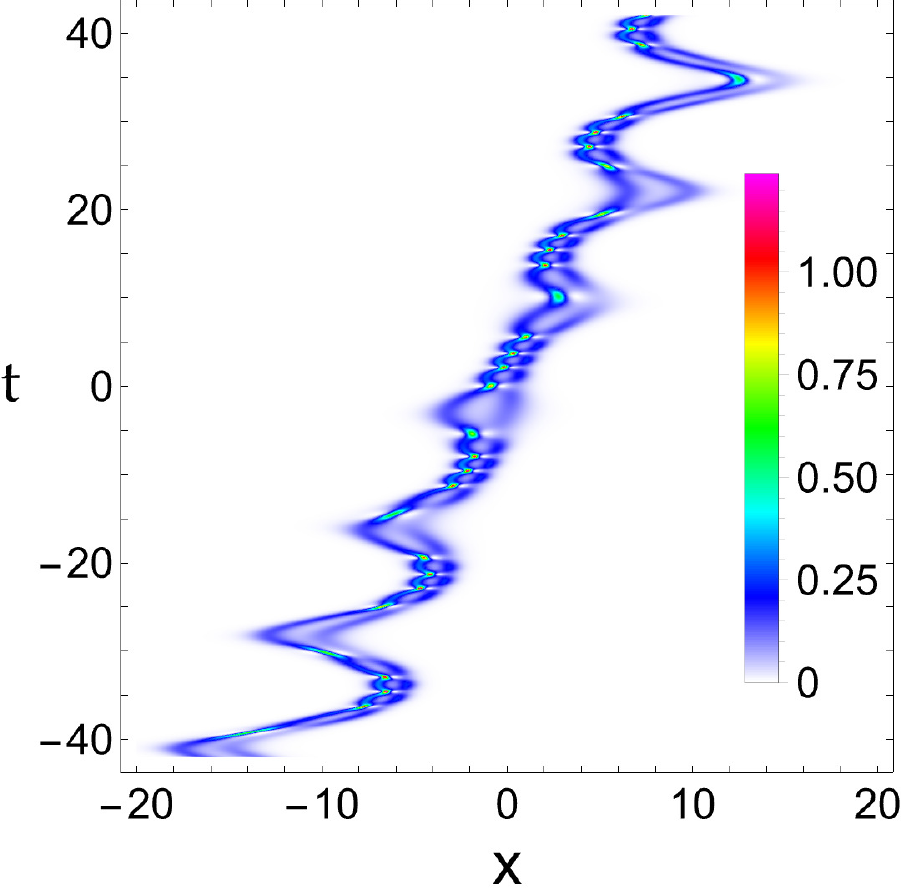}\qquad \includegraphics[width=0.34\linewidth]{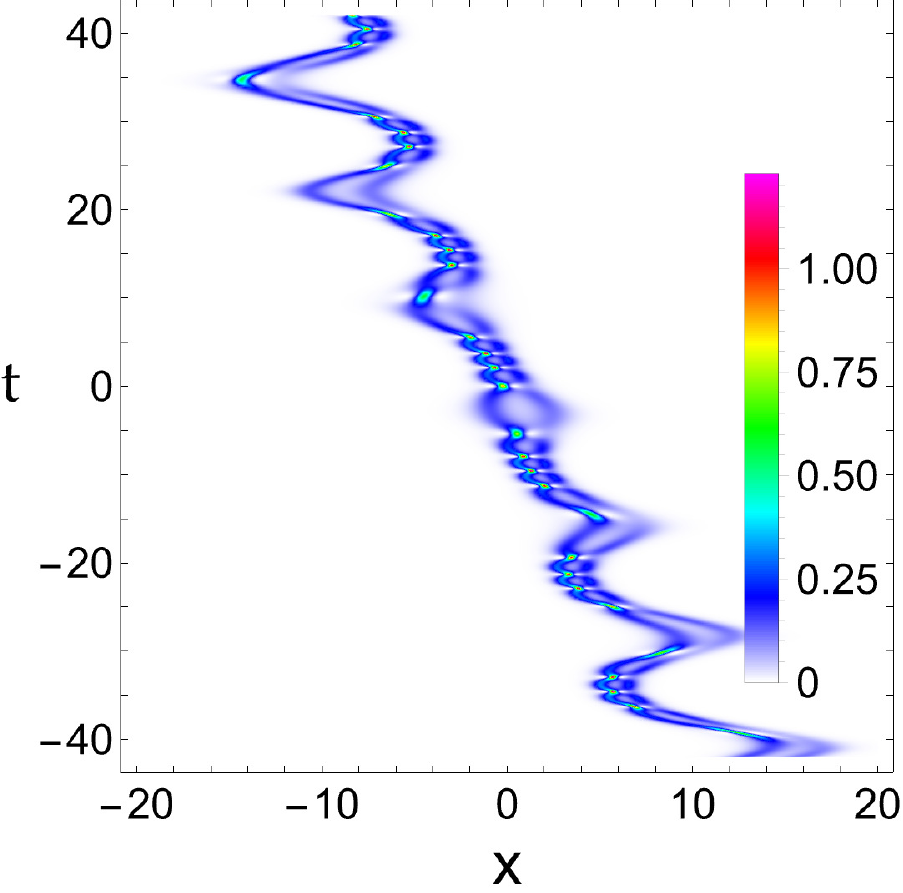}\\
		\includegraphics[width=0.34\linewidth]{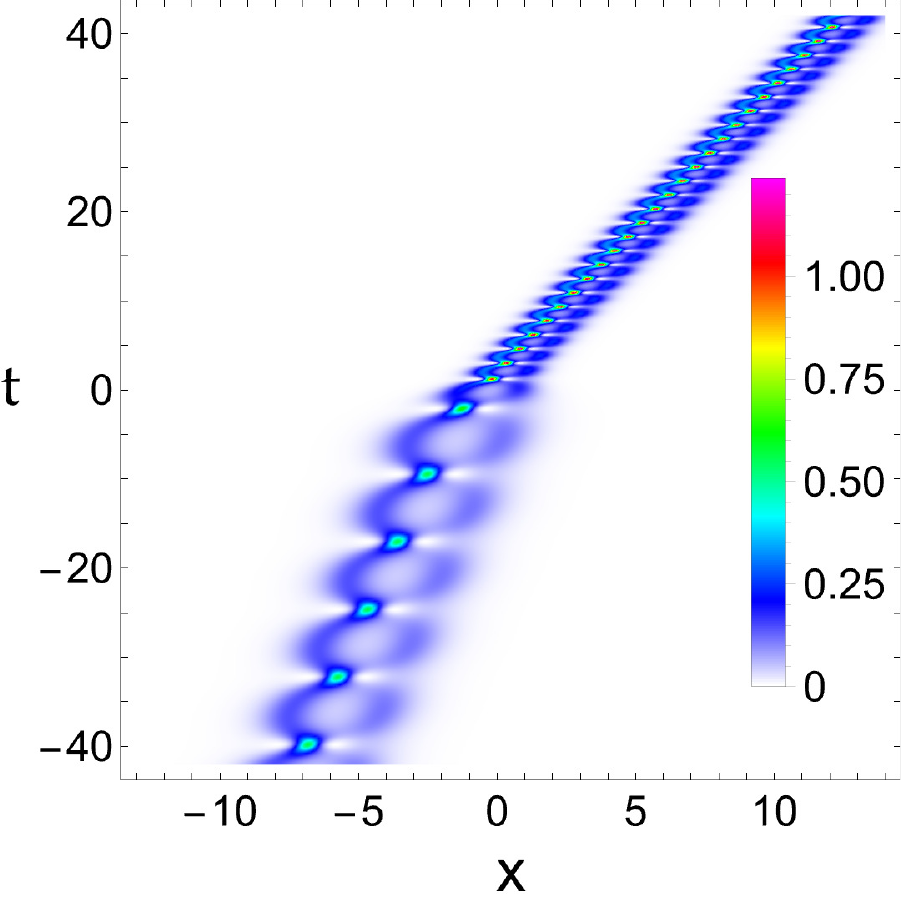}\qquad \includegraphics[width=0.34\linewidth]{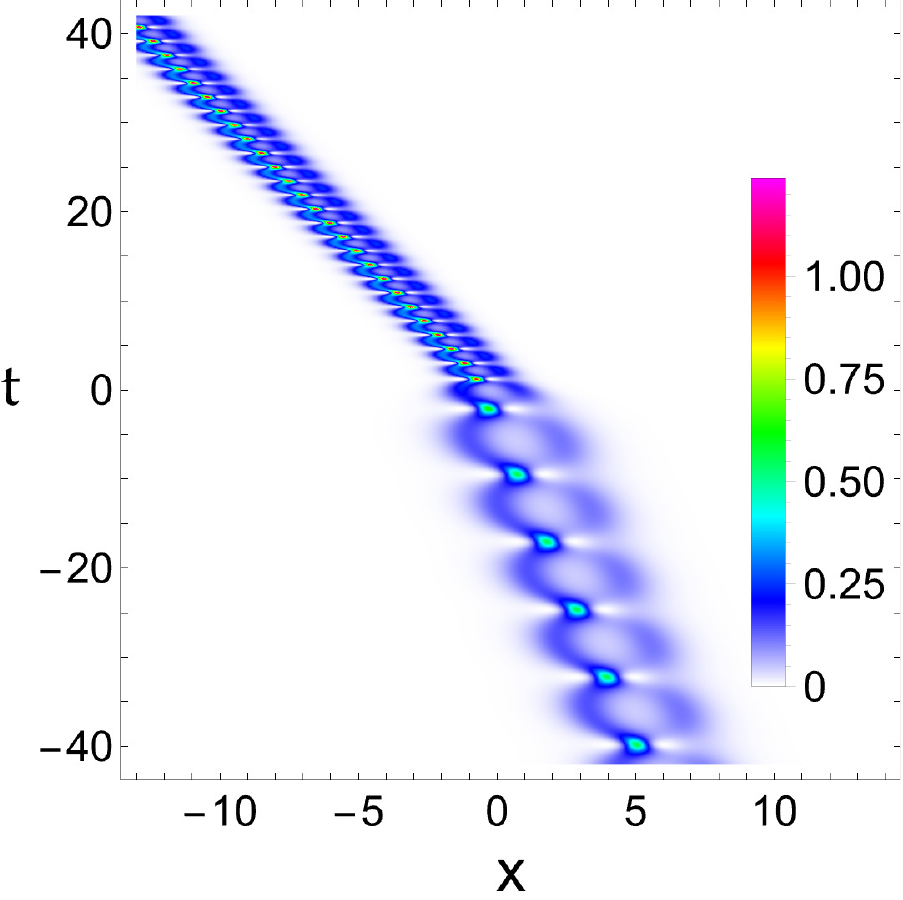}
		\caption{Non-autonomous FSM having degenerate type symmetric single-hump profile along all three components ($| \psi_1| =| \psi_0| =| \psi_{-1}| $) undergoes snaking oscillations due to periodic nonlinearity (top panel) and an amplification accompanied with compression and bending due to kink-like nonlinearity (bottom panel). The parameters are chosen as $\xi_1=0.52$, $p_1=2.0$, $p_2=0.75$, $p_3=0.5$, $p_4=0.0$, $k_1=1$, $k_2=1.7$,  $\alpha_1^{(1)}=\alpha_1^{(2)}=\alpha_1^{(3)}=0.2$, and   $\alpha_2^{(1)}=\alpha_2^{(2)}=\alpha_2^{(3)}=0.3$. The velocity of initial stationary FSM is controlled to become a right-moving (left-moving) bound molecule for $\xi_1=0.21$ ($\xi_1=0.21$).}
		\label{fig-nafsm-sym1}
	\end{figure}
	The above-considered periodic and localized-kink-like non-autonomous nonlinearities are fixed by taking $m=0$ and $m=1$ in Eq. (\ref{nonlin}), respectively, impacting the dynamics of observed FSM. Especially, the former sinusoidal nonlinearity induces a periodically oscillating (snaking) pattern for FSM, where the already breathing structure undergoes further oscillations in both amplitude and position. Also, it increases the asymmetry between the two breathing solitons and the period of oscillations. As pointed out already, the localization and velocity of the molecule can be manipulated by $\xi_1$ and $\xi_2$. Next, due to a tan-hyperbolic nonlinearity function, the breathing molecule executes an amplification associated with compression and increased periodic oscillations. Simply, we can observe a step-like operation where the amplitude and period of breathing oscillations are less before the step function. However, after implementing step nonlinearity, the amplitude and period of breathing oscillations undergo a step increase, which reduces the width of the breathing molecule and is nothing but a compression. Further, the period of oscillation of the soliton chain or molecule is increasing due to this kink/step-like nonlinearity modulation, which developed more number of breathing chains/patterns. 
	Another noticeable effect is the bending of the soliton molecule. This is a consequence of the varying velocity of the molecule around the interaction regime ($x=t=0$) due to the inhomogeneity of the nonlinearity. Ultimately, this influences the width and central position of the soliton molecule. The sense of bending can be altered by varying the $\xi_1$ and $\xi_2$ parameters as evidenced from Fig. \ref{fig-nafsm-sym1}. 
	For an easy understanding of the above phenomena, we have shown the respective graphical demonstration for a periodic oscillation and amplification/compression of an initial stationary FSM in Fig. \ref{fig-nafsm-sym1}. Here, we can find that the stationary (zero-velocity) nature of the molecule is now transformed into a moving bound structure due to the controllable similarity variables $\xi_1$ and $\xi_2$. Also, one can consider a travelling molecule and study the modulational characteristics of non-autonomous nonlinearities.
	
\begin{figure}
		\centering\includegraphics[width=0.3\linewidth]{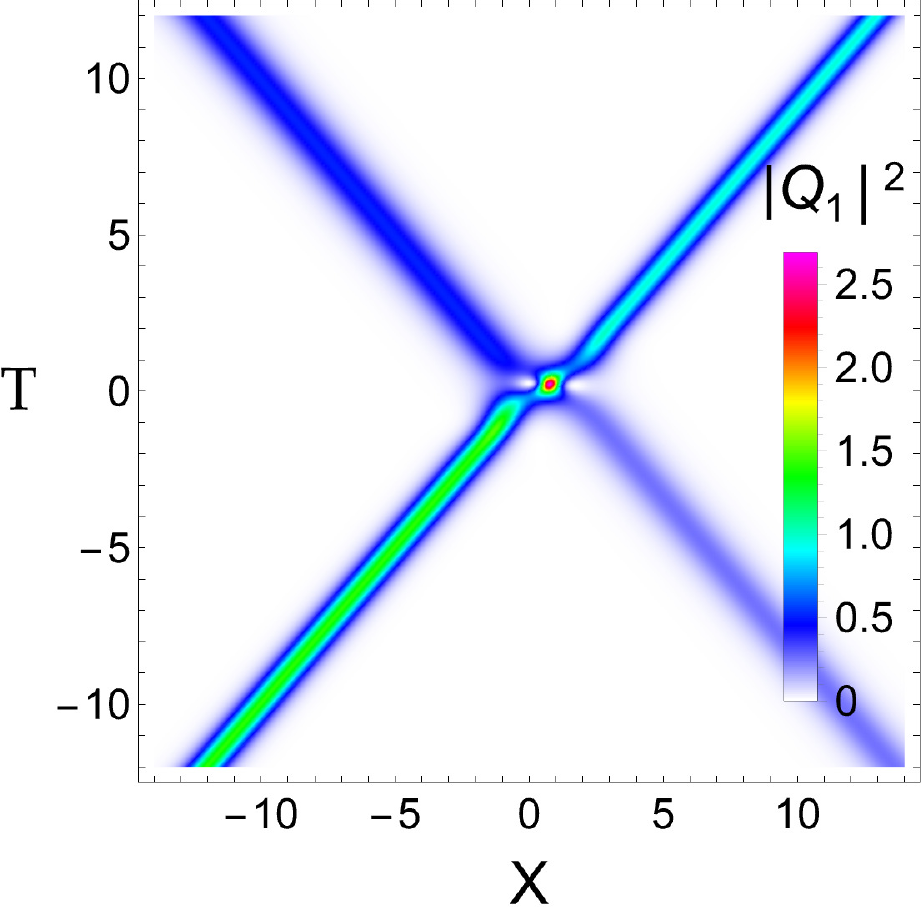}\quad \includegraphics[width=0.3\linewidth]{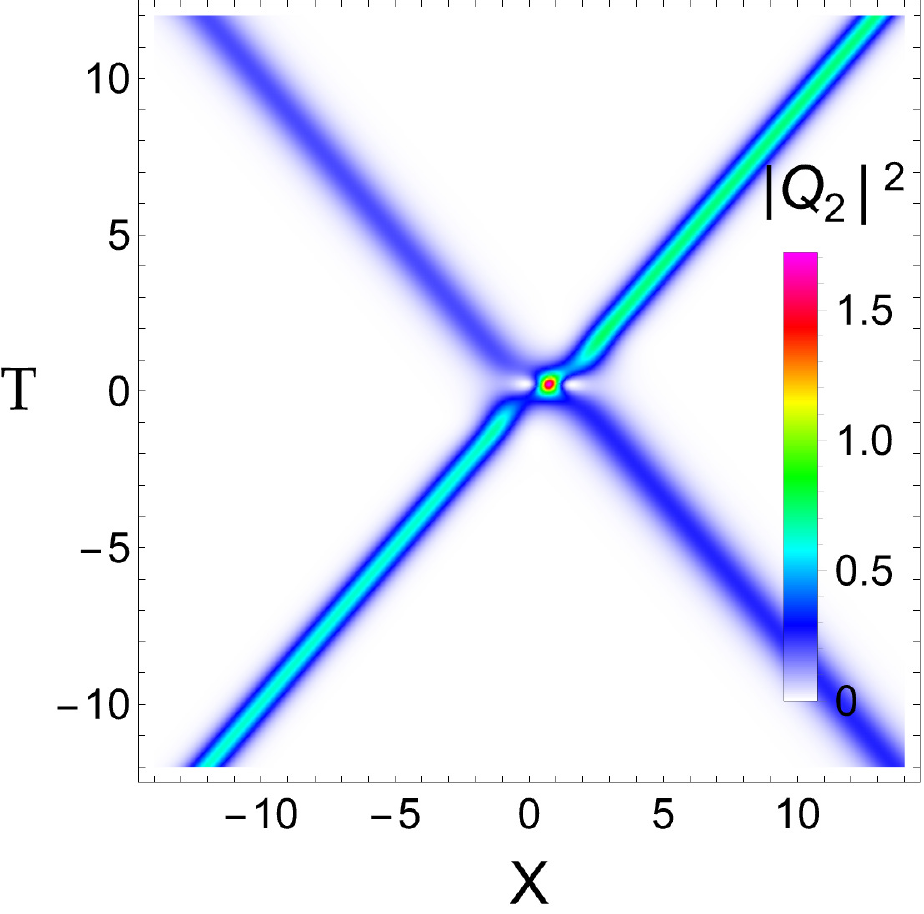}\quad
		\includegraphics[width=0.3\linewidth]{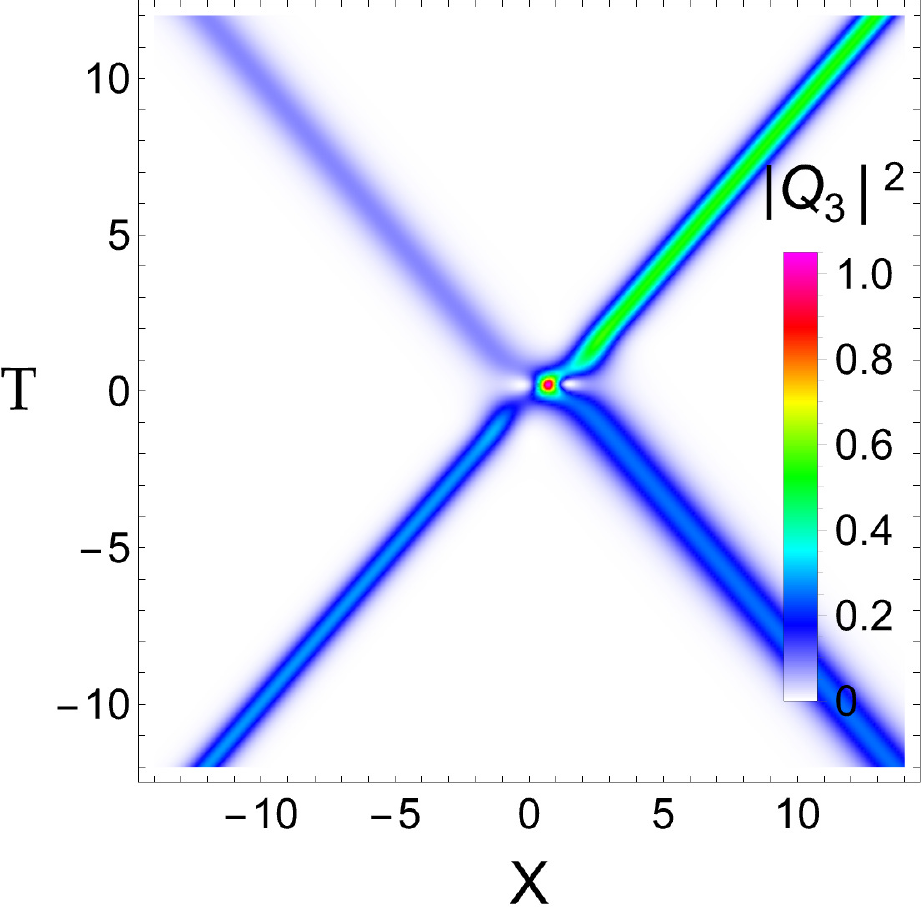}\\
		\centering\includegraphics[width=0.3\linewidth]{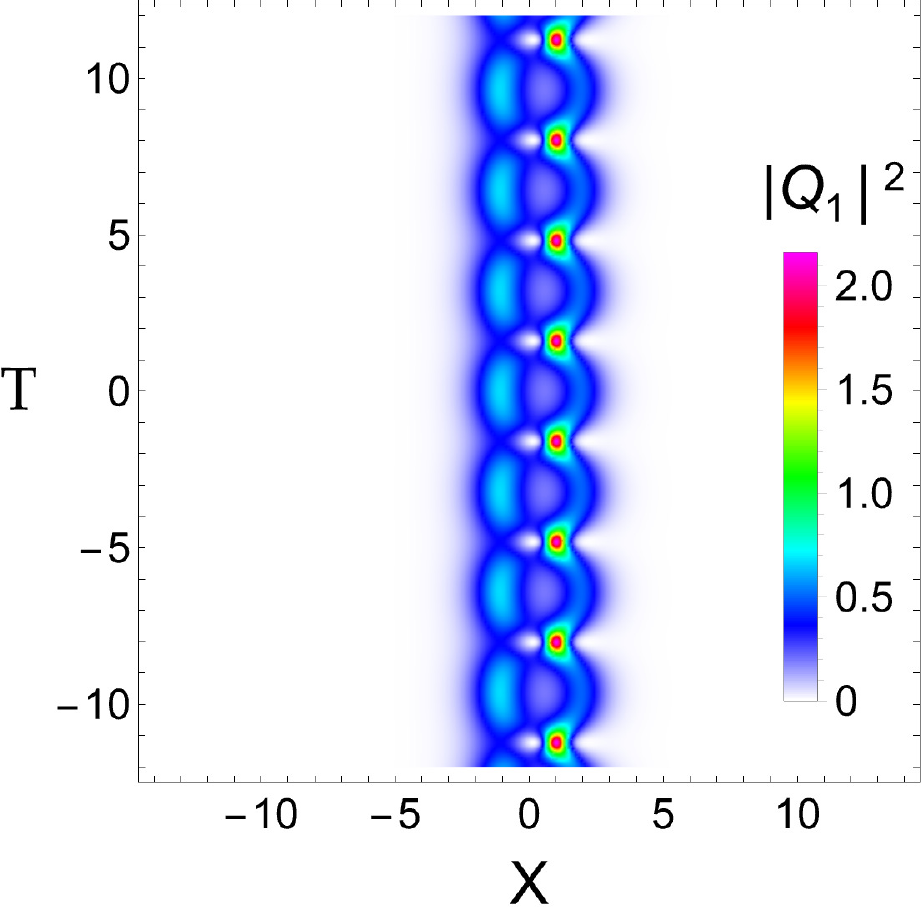}\quad \includegraphics[width=0.3\linewidth]{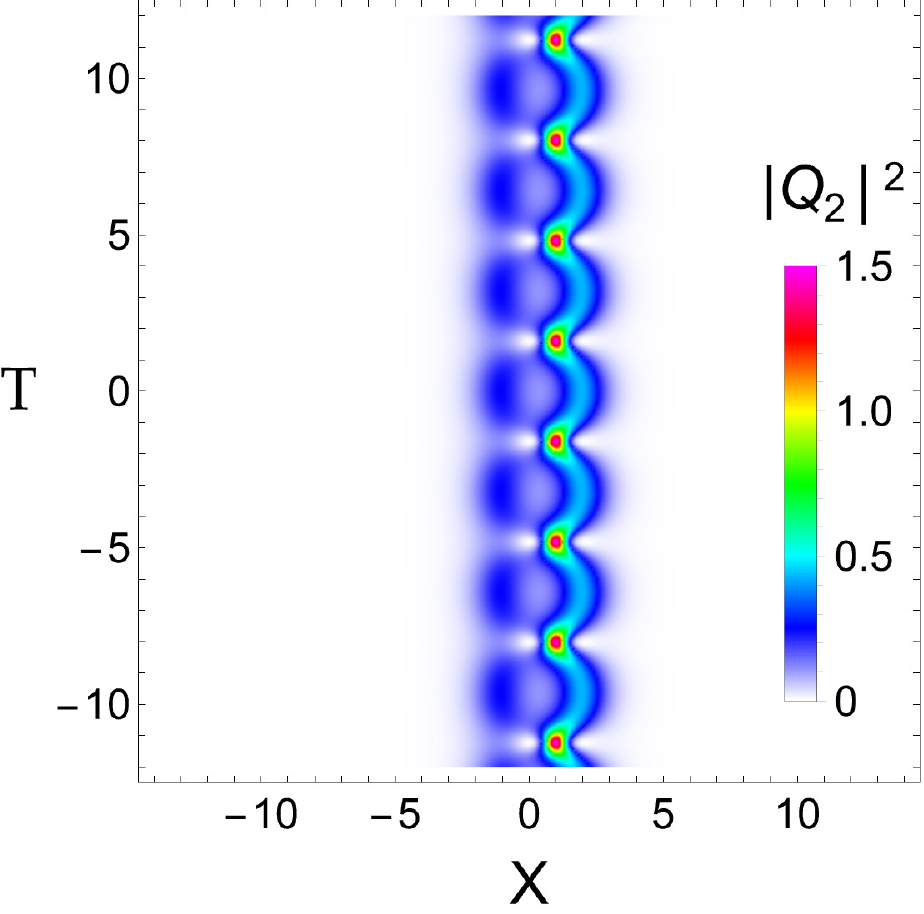}\quad
		\includegraphics[width=0.3\linewidth]{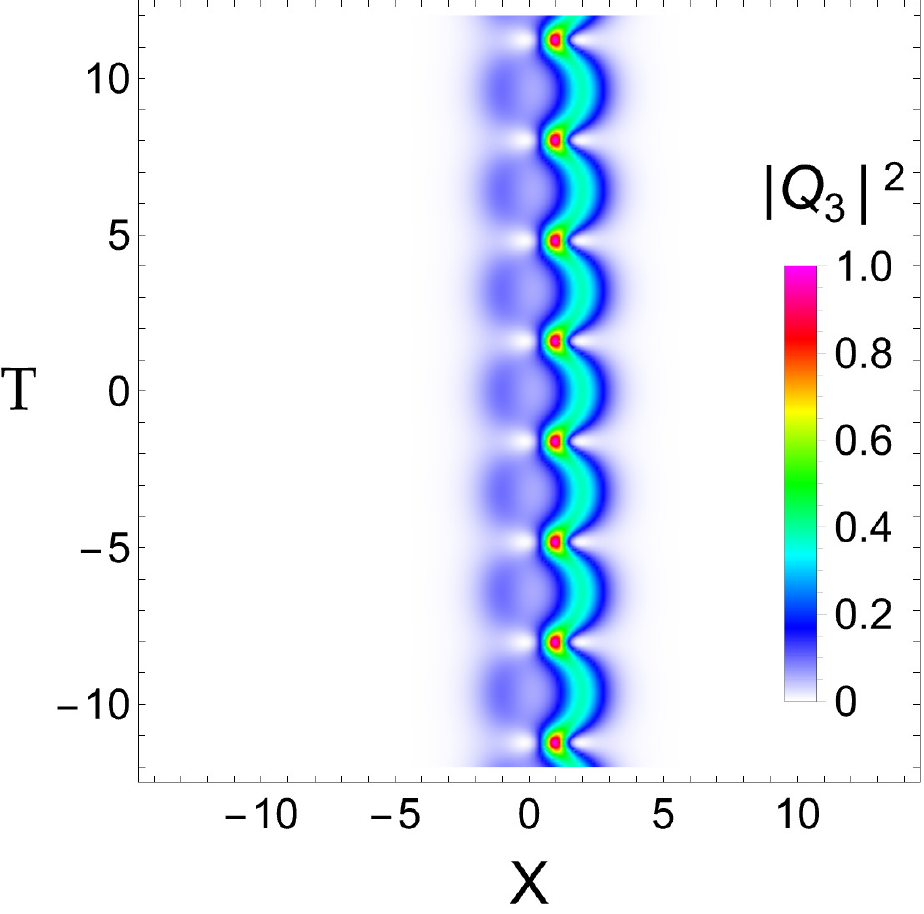}\\
		\centering\includegraphics[width=0.3\linewidth]{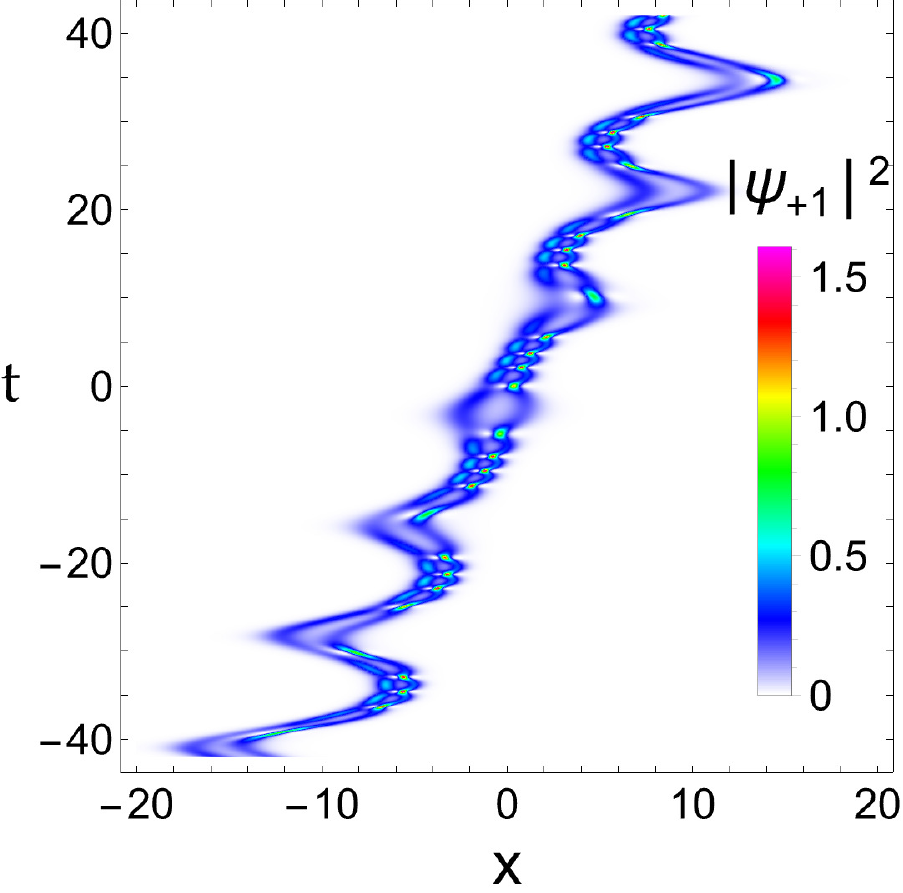}\quad \includegraphics[width=0.3\linewidth]{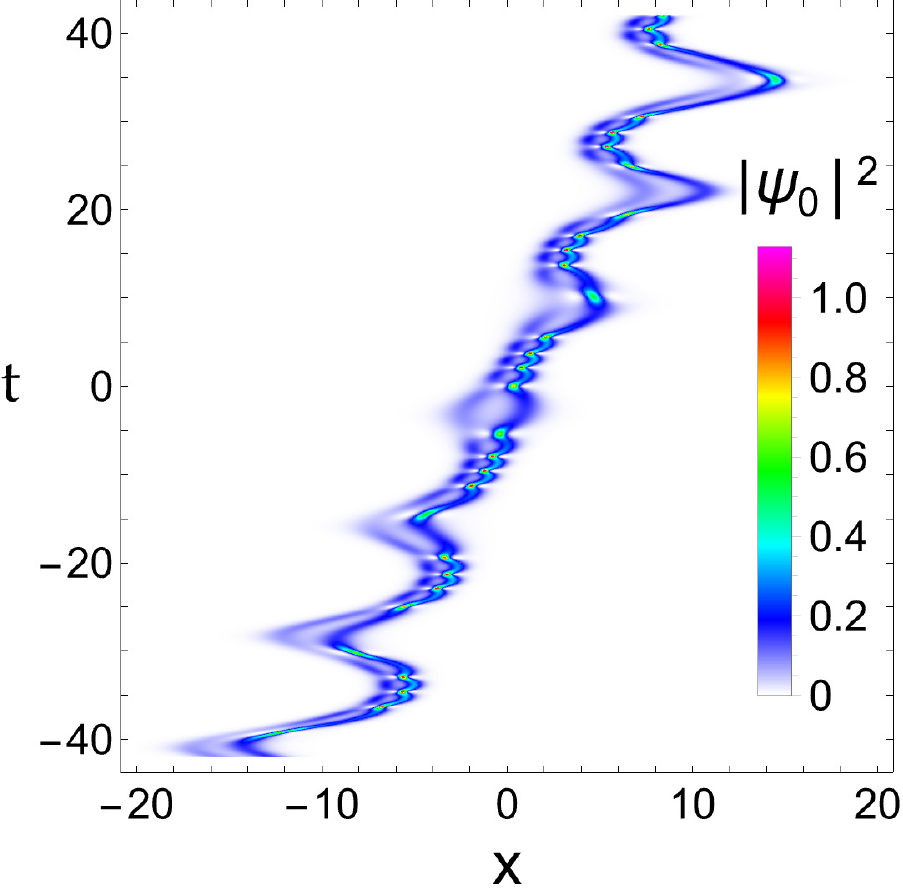}\quad 
		\includegraphics[width=0.3\linewidth]{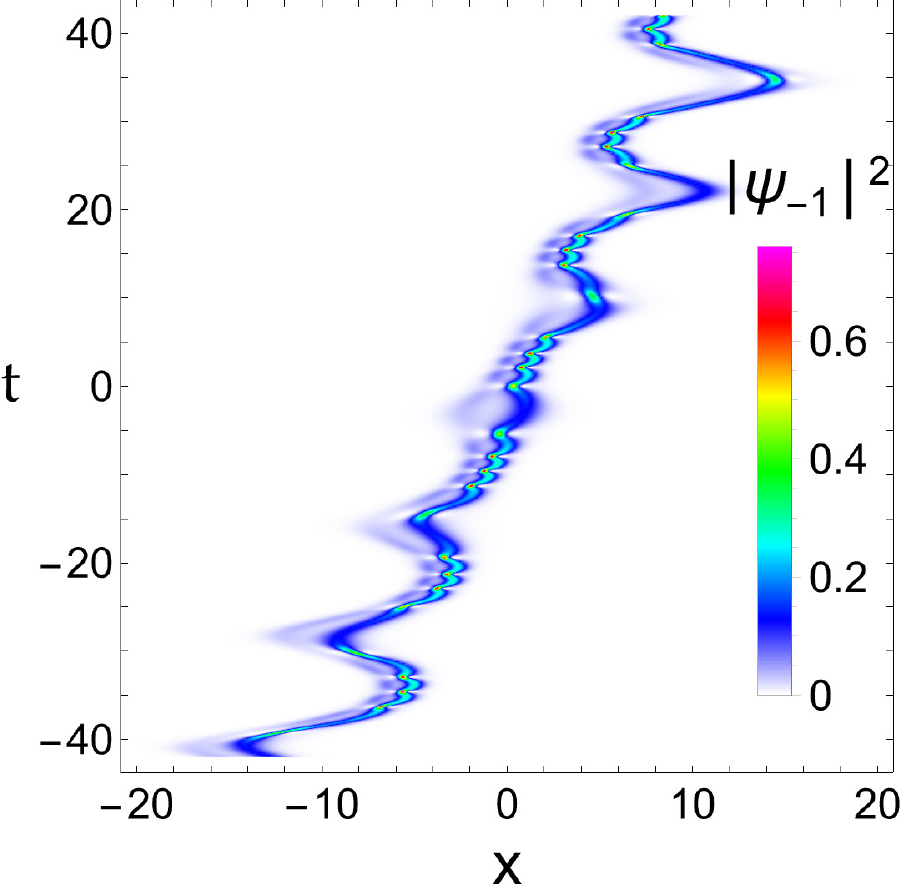}\\
		\centering\includegraphics[width=0.3\linewidth]{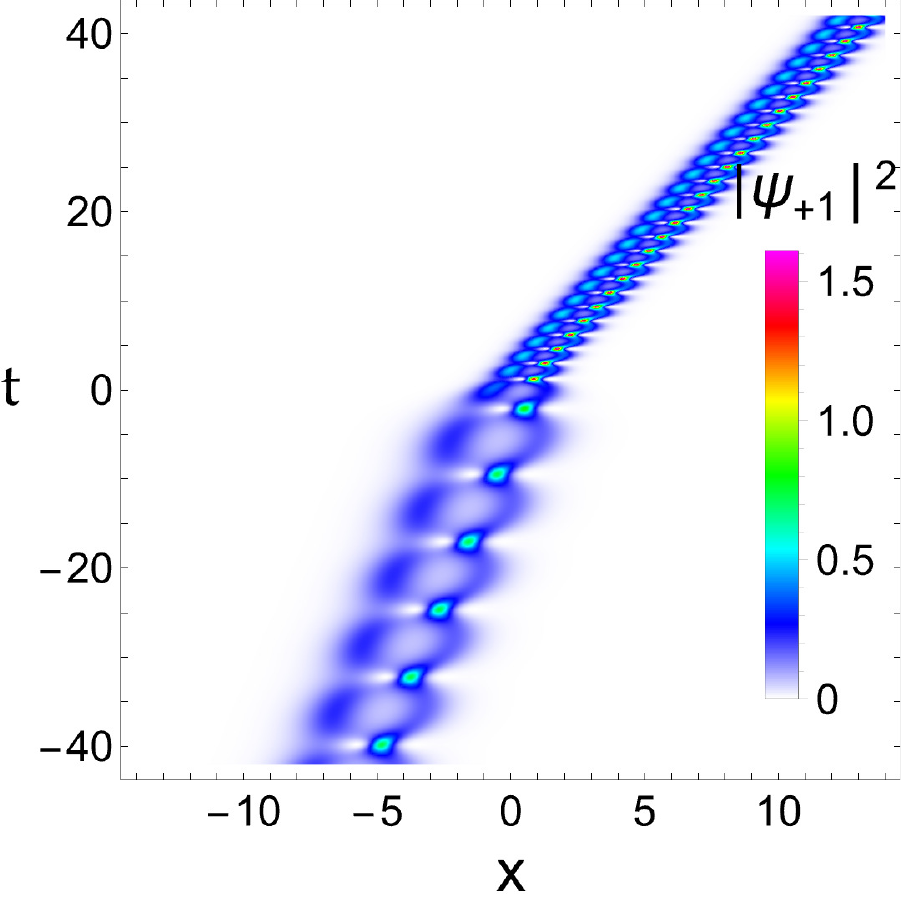}\quad \includegraphics[width=0.3\linewidth]{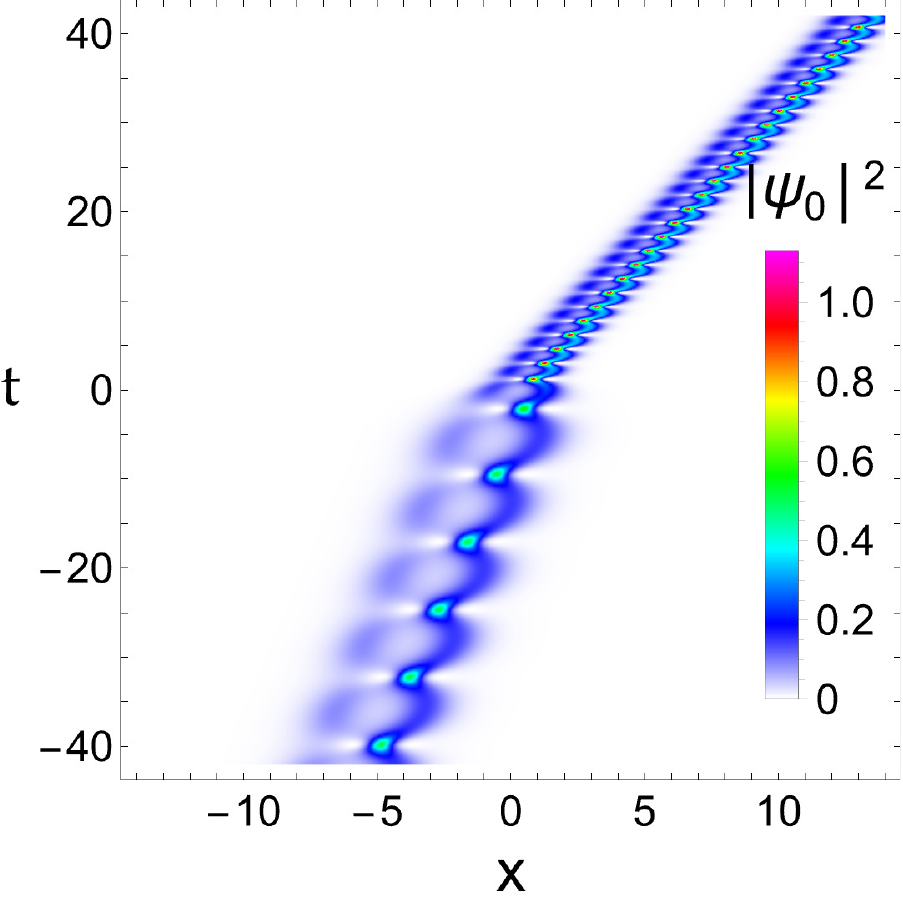}\quad 
		\includegraphics[width=0.3\linewidth]{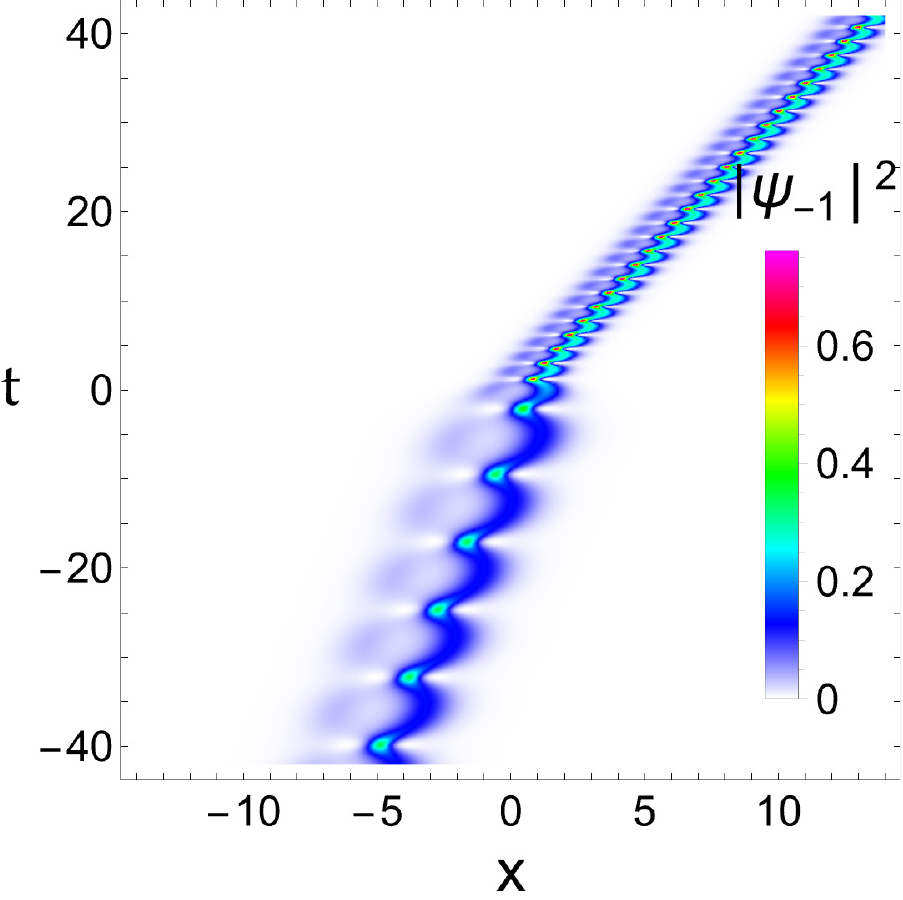}
		\caption{Non-degenerate autonomous FSs ($|Q_1|\neq |Q_2| \neq |Q_3|$) undergoing inelastic interaction (first row: $k_1=1+0.5i$ \& $k_2=1.7-0.5i$) for the choice $\alpha_1^{(1)}=\alpha_1^{(2)}=\alpha_1^{(3)}=0.2$ and $\alpha_2^{(1)}=2.25,\alpha_2^{(2)}=1.5,\alpha_2^{(3)}=1.0$ and their transition to stationary FSM (second row: $k_1=1+0.0i$ \& $k_2=1.7+0.0i$). Non-autonomous right-moving FSM is influenced by a periodic nonlinearity (third row) and kink-like nonlinearity (fourth row) for $\xi_1=0.52$ and $\xi_2=0.21$.}
		\label{fig-nafsm-asym}
	\end{figure}

	The above-discussed dynamics correspond to a degenerate type FSM ($|Q_1|=|Q_2|=|Q_3|$). 
	However, we can also have another possibility of non-degenerate FSM formed out from two bright solitons with distinct densities in each component ($|Q_1|\neq |Q_2| \neq |Q_3|$) by choosing the parameters $\alpha^{(1)}_u\neq \alpha^{(2)}_u \neq \alpha^{(3)}_u$, $u=1,2$. This choice, along with different velocities, usually results in an inelastic (spin-precession) type interaction of two solitons as it leads to $\Gamma_1=\Gamma_2=0$ but $\Gamma_3\neq 0$. We have shown such inelastic interaction of two non-degenerate FSs travelling with opposite velocities ($k_{1I}>0>k_{2I}$) in the top panel of Fig. \ref{fig-nafsm-asym} (for an easy reference). 
	Interestingly, during molecule formation, they execute parallel-type propagation for coinciding central positions due to non-degenerate peaks instead of single-peak-soliton. Further, we have demonstrated its transformation into a stationary molecule with a breathing pattern through the velocity matching ($k_{1I}=0=k_{2I}$) in the second row of Fig. \ref{fig-nafsm-asym}. The influence of non-autonomous (periodic and kink-like) nonlinearities in the third and fourth rows of  Fig. \ref{fig-nafsm-asym} demonstrates the asymmetric oscillation and amplification-cum-compression of FSM.
	
	\subsection{\bf Polar Soliton Molecule (PSM)}
	Next, the polar soliton molecule (PSM) can be formed from the interaction solution of two PSs. To be precise, the general two-soliton solution (\ref{2sol}) and (\ref{nonat2sol}) correspond to autonomous and non-autonomous PSs without any restriction on spin polarization parameters $\alpha_u^{(j)},~u=1,2,~j=1,2,3,$ that can take arbitrary complex values and keep $\Gamma_j\neq 0,~j=1,2,3$. In this case, every term arising in the general solution (given in the Appendix) becomes non-vanishing and contributes to the dynamics of PSs and thereby, PSM can be constructed with velocity resonance condition. Due to the presence of every term in the PSM, it is difficult to write the explicit form of such PSM for non-coinciding central positions $k_{1R} \neq k_{2R}$ in a compact manner. However, for a special case with coinciding central position $k_{1R} = k_{2R}$, we can rewrite the PSM in terms of trigonometric and hyperbolic functions as below.
	\bes \bea
	&&Q_j(X,T)=\frac{A_j~ \mbox{cosh}\left(\eta_{1R}+N_R^{(j)}+i N_I^{(j)}\right) e^{i\eta_{1I}}}{ e^{\frac{\chi_3}{2}} \mbox{cosh}\left(2\eta_{1R}+\frac{\chi_3}{2}\right)+ e^{\chi_4}}, \quad  j=1,2,3,\label{apsm}\\
	&&\left(\psi_{+1},~ \psi_{0},~\psi_{-1}\right)^T = \xi_1\sqrt{{c}(t)}~ (Q_1,~Q_2,~Q_3)^T~ e^{i\theta(x,t)},\qquad j=1,2,3,\label{napsm}
	\eea \ees 
	where $A_j=\exp{\left(\frac{\chi_2^{(4-j)}+\chi_1^{(4-j)}}{2}\right)}$, $N_R^{(j)}= \frac{\chi_{2R}^{(4-j)}-\chi_{1R}^{(4-j)}}{2}$, and $N_I^{(j)}= \frac{\chi_{2I}^{(4-j)}-\chi_{1I}^{(4-j)}}{2}$ with $\exp({\chi_1^{(j)}})=\alpha_1^{(j)}+\alpha_2^{(j)}$, $\exp({\chi_2^{(j)}})=\sum_{u,v=1}^2 (e^{\delta_{uv}^{(j)}}+e^{\delta_{u}^{(j)}})$, $\exp({\chi_3})=R_3+\sum_{u,v=1}^2 (e^{\epsilon_{uv}}+e^{\tau_{u}}+e^{\tau_{u}^*})$, $\exp({\chi_4})=(e^{R_1}+e^{R_2}+e^{\delta_0}+e^{\delta_0^*})/2$ with $k_2=k_1$ and other relations takes the form as given in Eq. (\ref{str}) and in \ref{appendix}.\\ 
	
\begin{figure}
		\centering\includegraphics[width=0.3\linewidth]{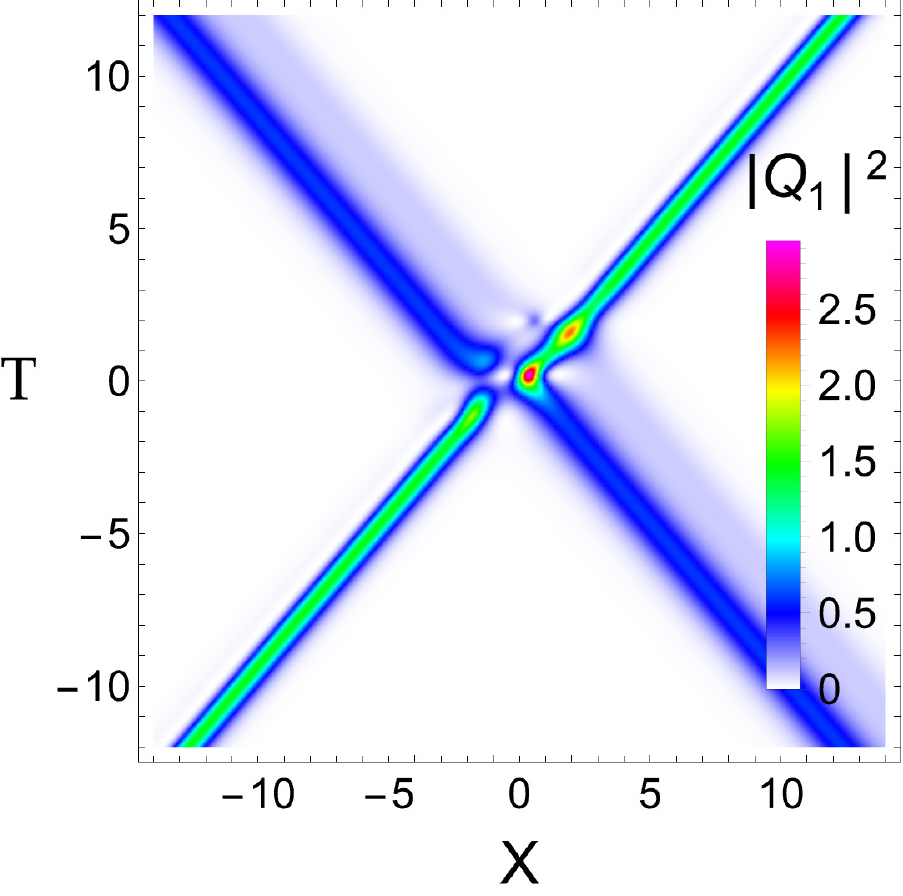}\quad \includegraphics[width=0.3\linewidth]{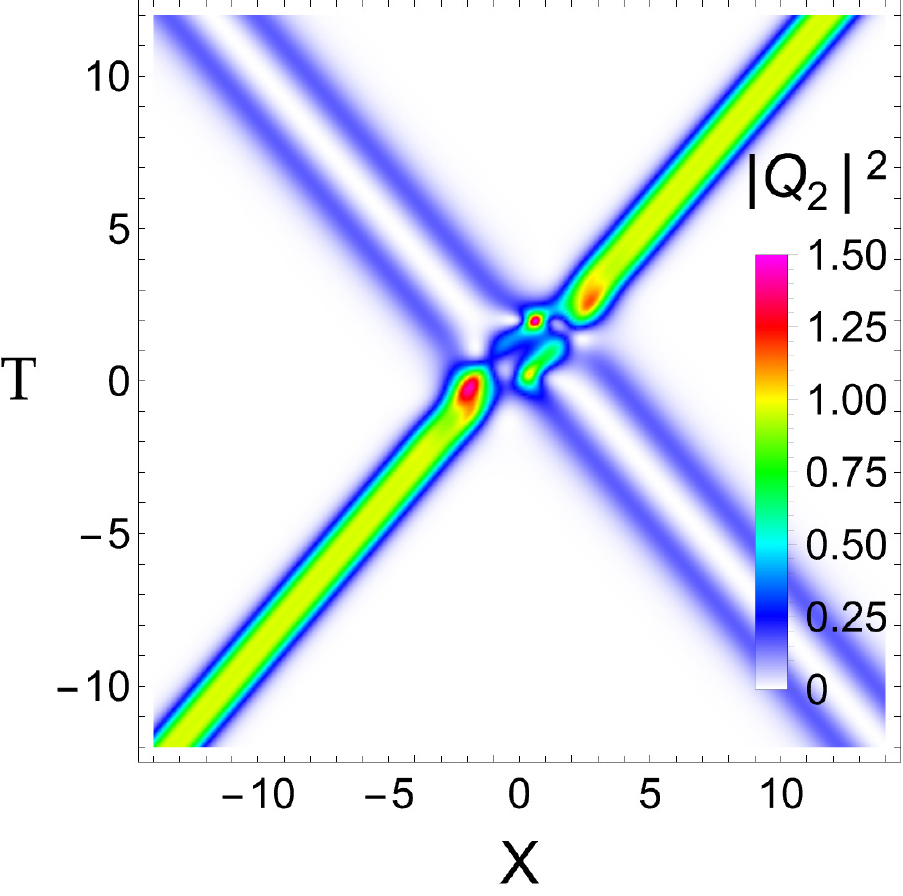}\quad
		\includegraphics[width=0.3\linewidth]{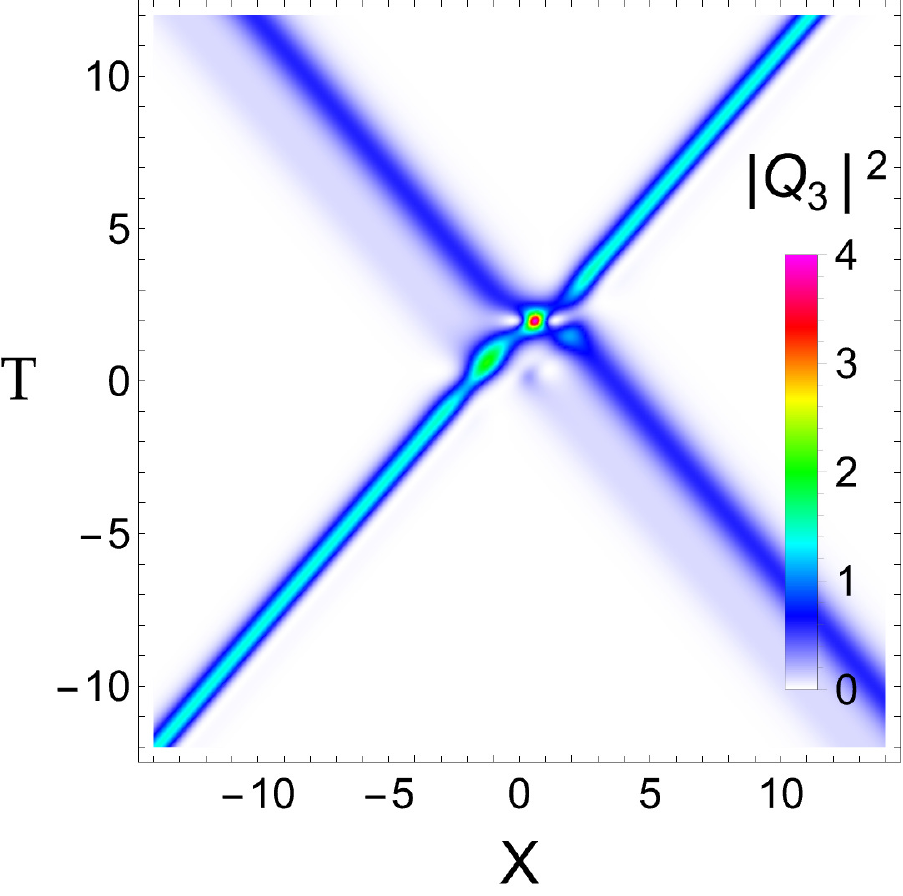}\\
		\centering\includegraphics[width=0.3\linewidth]{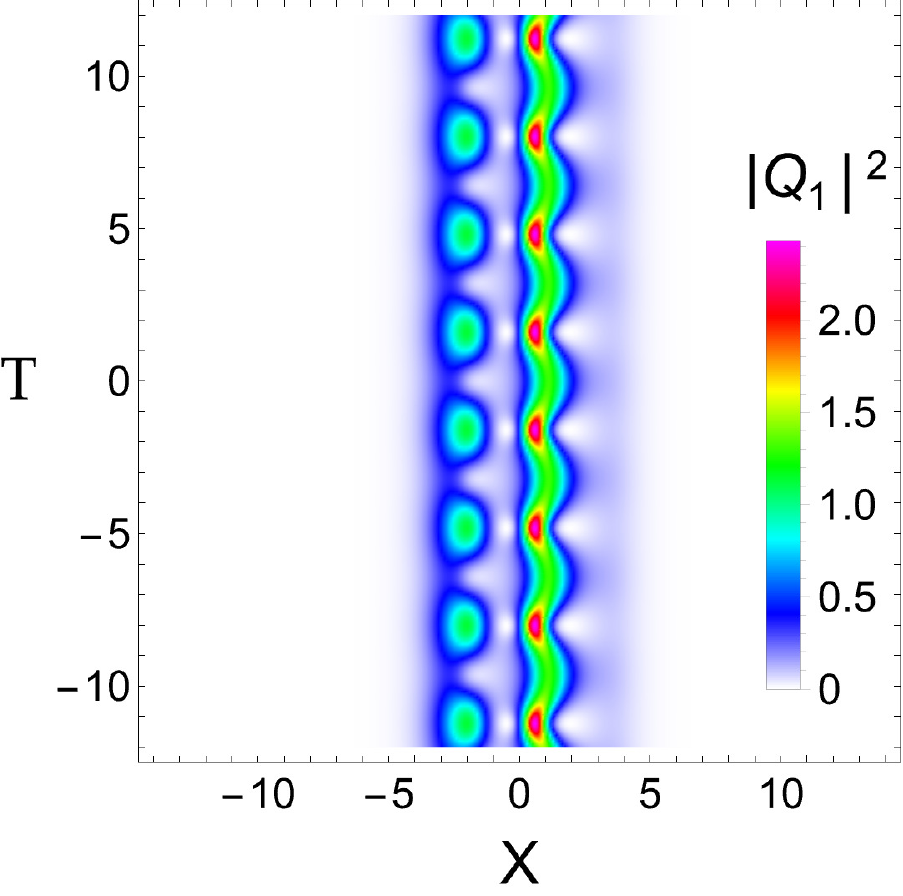}\quad \includegraphics[width=0.3\linewidth]{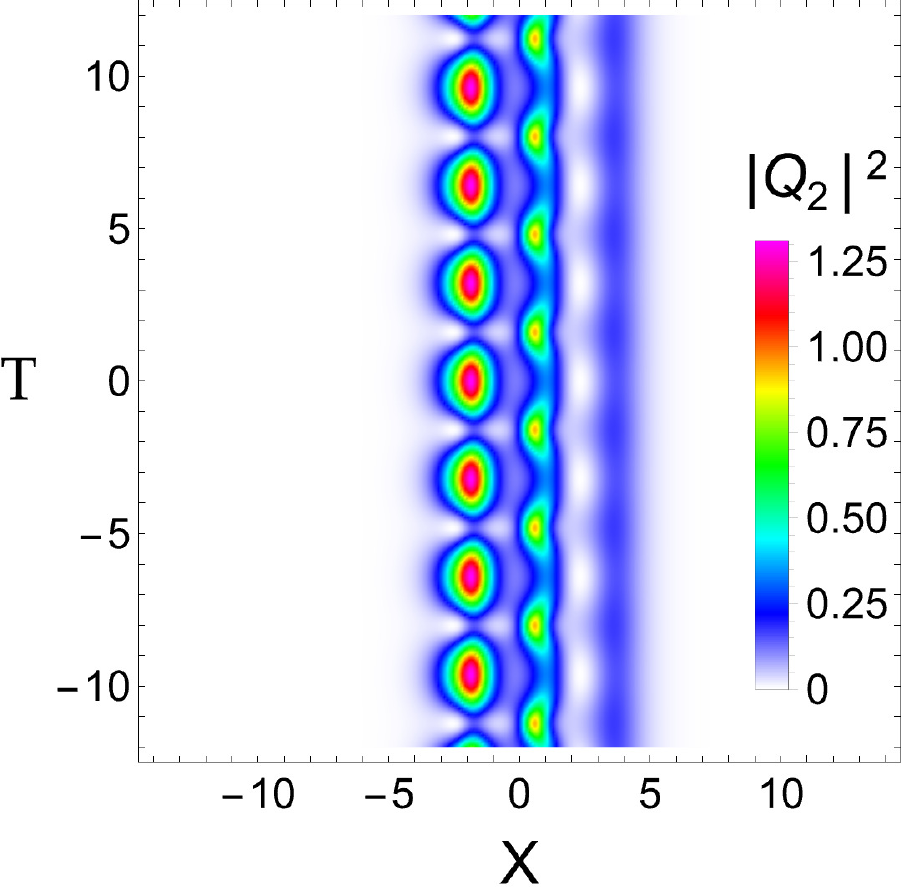}\quad
		\includegraphics[width=0.3\linewidth]{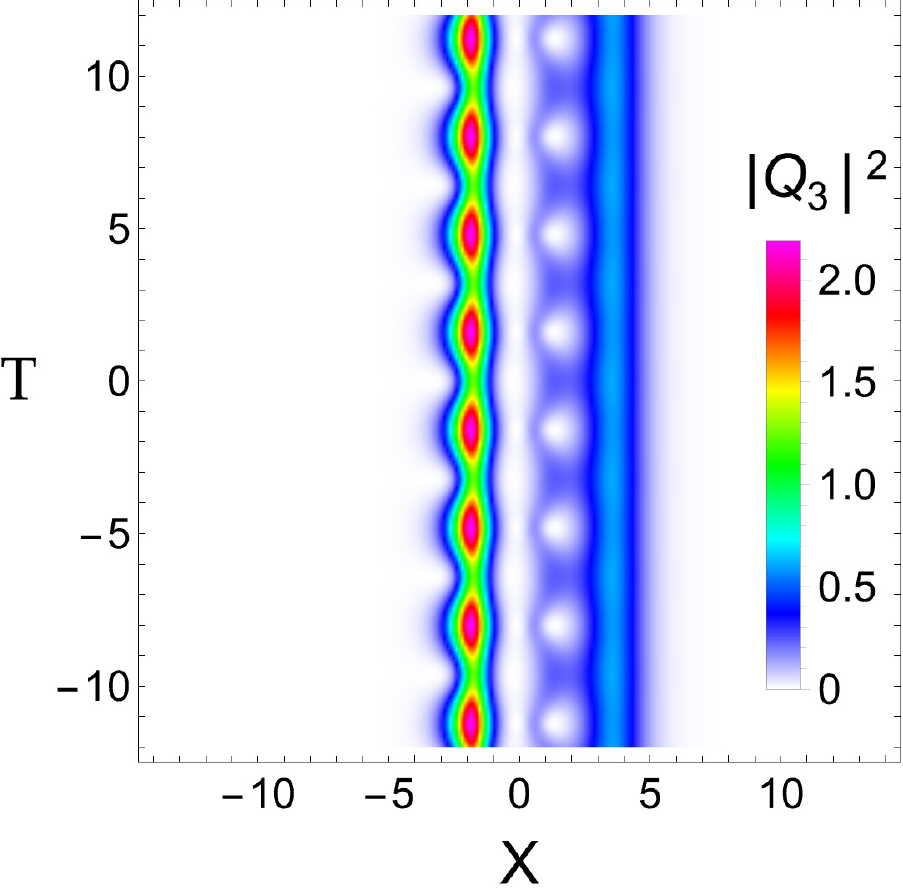}\\
		\centering\includegraphics[width=0.3\linewidth]{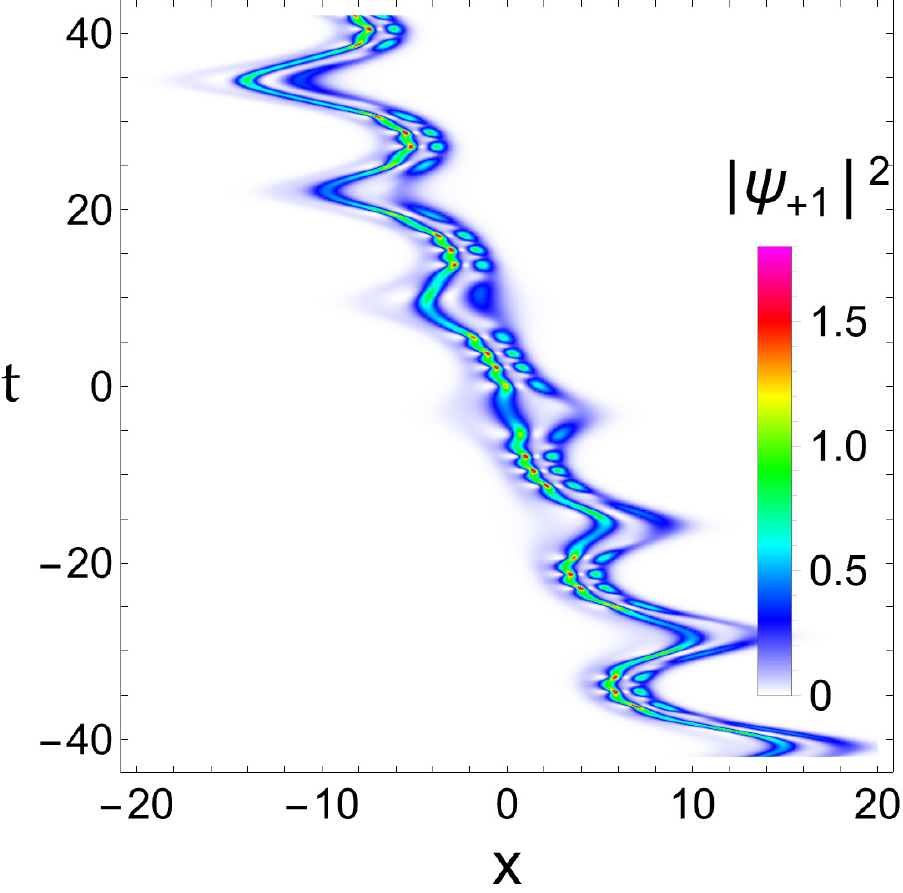}\quad \includegraphics[width=0.3\linewidth]{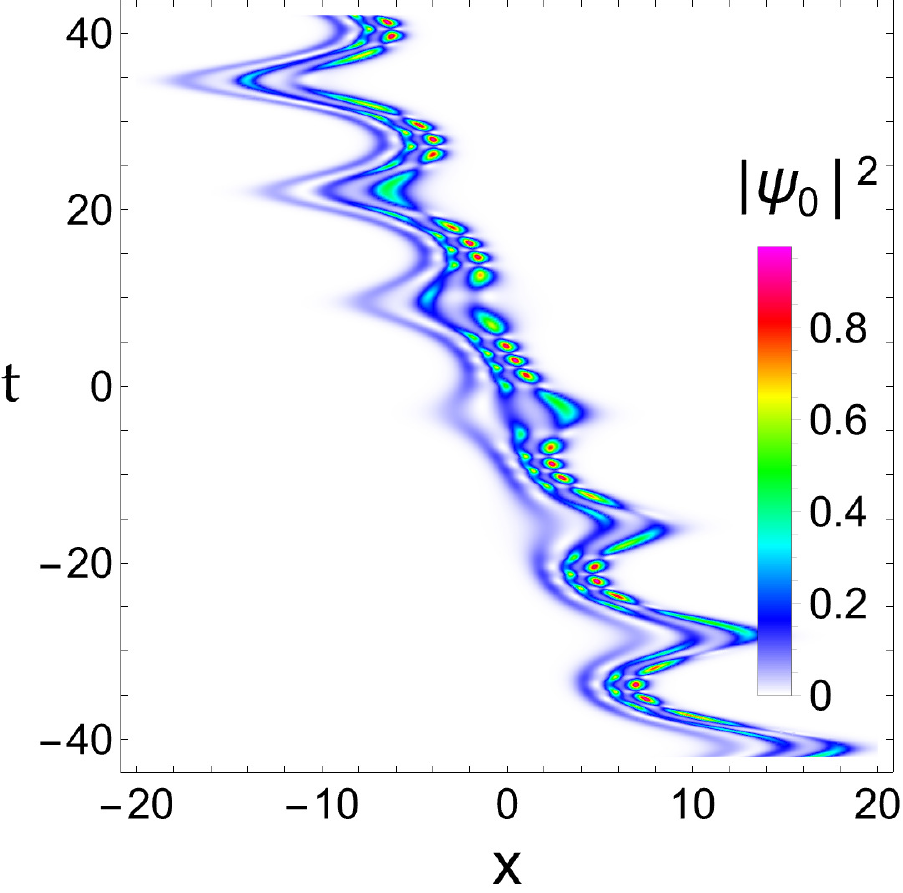}\quad 
		\includegraphics[width=0.3\linewidth]{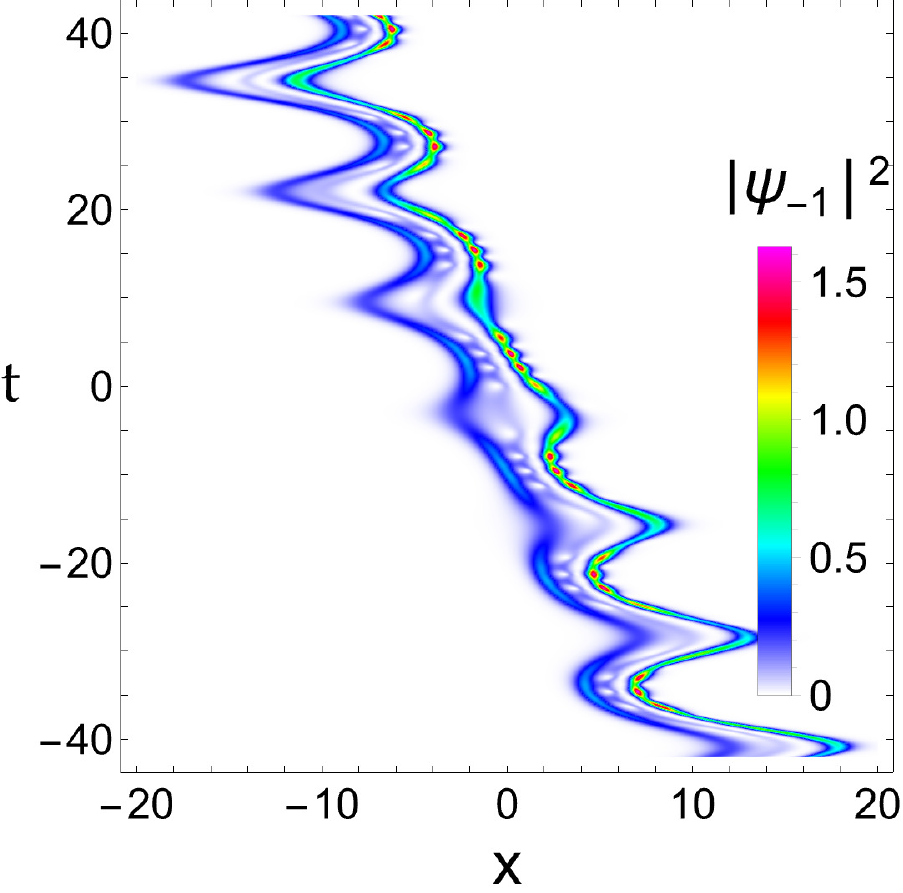}\\
		\centering\includegraphics[width=0.3\linewidth]{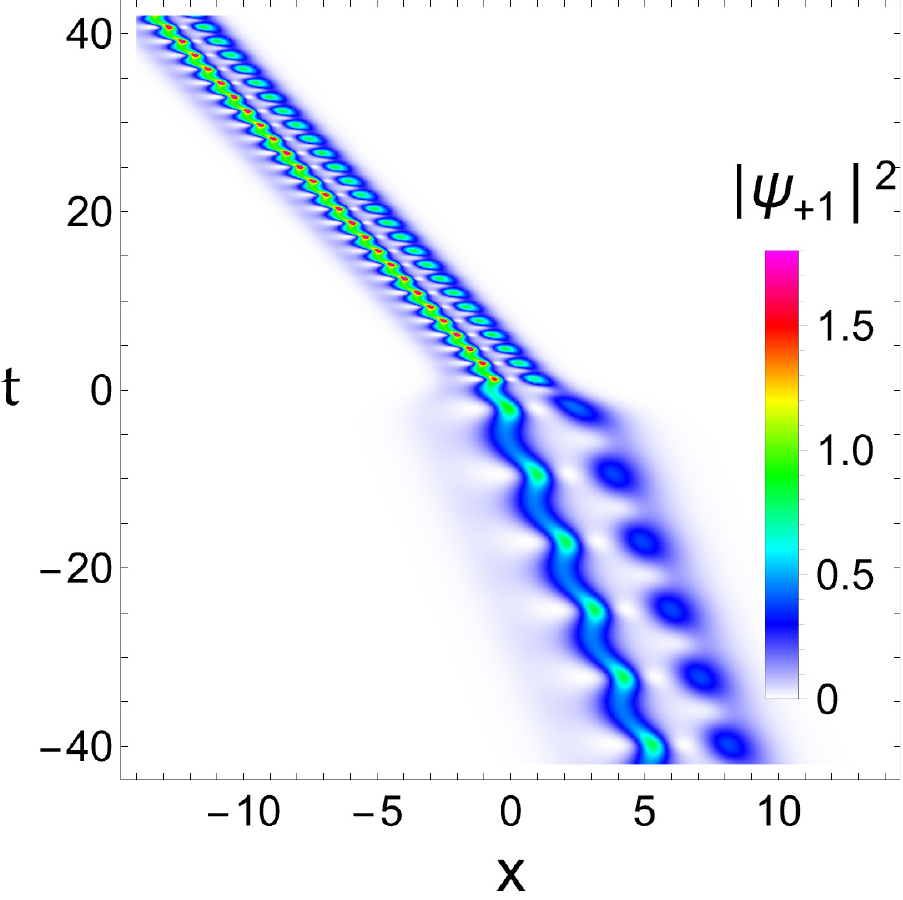}\quad \includegraphics[width=0.3\linewidth]{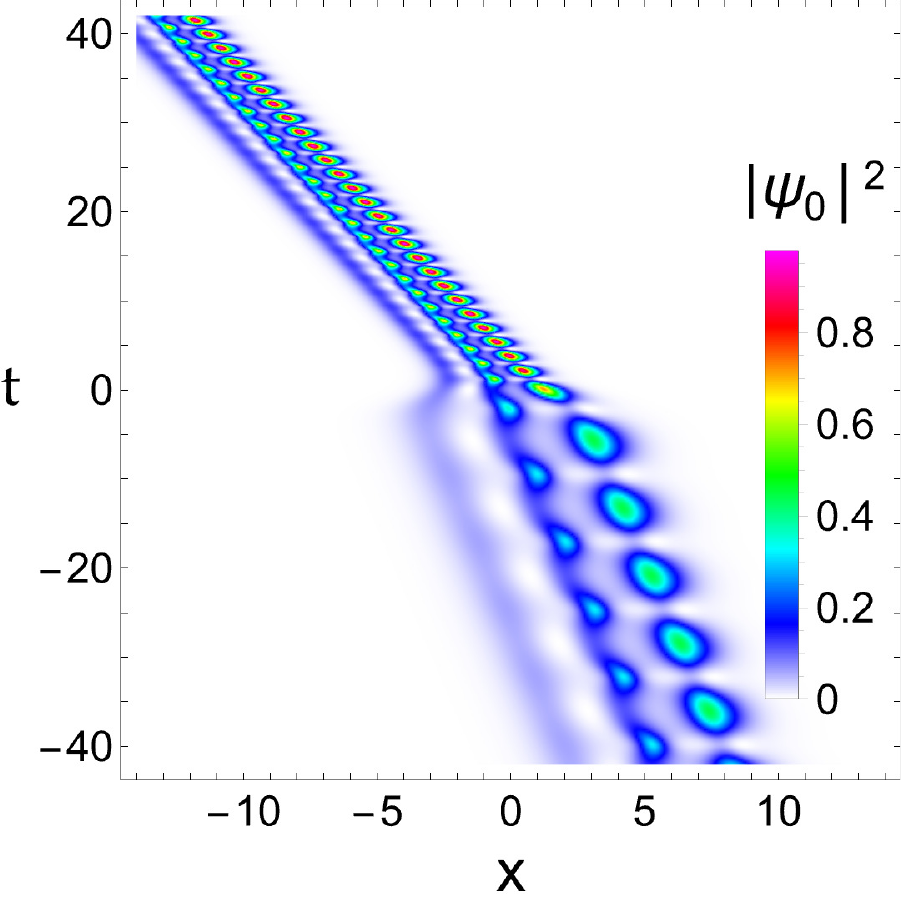}\quad 
		\includegraphics[width=0.3\linewidth]{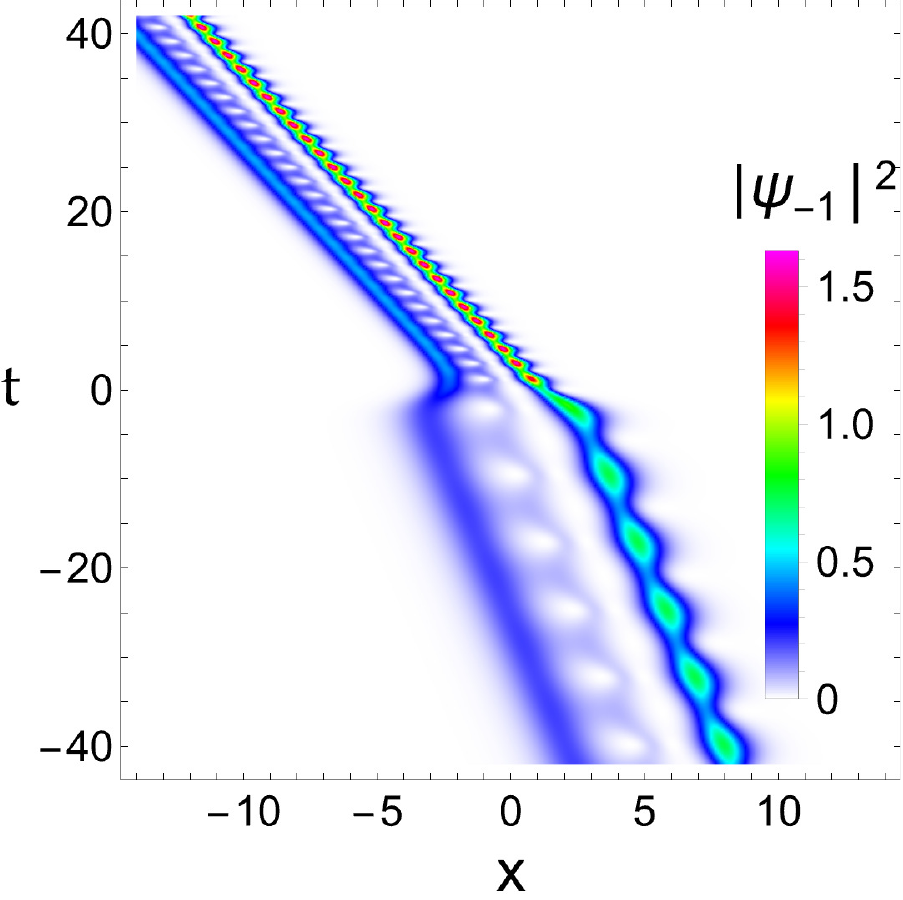}
		\caption{Elastic interactions of PSs ($|Q_1|$ \& $|Q_3|$: asymmetric-double-hump$\times$single-hump $|Q_2|$: double-hump$\times$flat-top) for $k_1=1+0.5i$ \& $k_2=1.7-0.5i$ (first row) and their transition to stationary PSM (second row: $k_1=1+0.0i$ \& $k_2=1.7+0.0i$). Non-autonomous PSM influenced by a periodic nonlinearity exhibiting a left-moving snaking effect (third row) and kink-like nonlinearity revealing a left-moving bending molecule with amplification-cum-compression (fourth row). Here the spin-polarization parameters are taken as $\alpha_1^{(1)}=1.1,~\alpha_1^{(2)}=0.6,~\alpha_1^{(3)}=0.4$ and $\alpha_2^{(1)}=0.75,\alpha_2^{(2)}=0.5$ and $\alpha_2^{(3)}=0.125$ with $\xi_1=0.52$ and $\xi_2=-0.21$.}
		\label{fig-napsm}
	\end{figure}

	The above PSM solution is nothing but two identical PSs ($\eta_2=\eta_1$) driven by six spin-polarization parameters $\alpha_u^{(j)},~u=1,2,~j=1,2,3$. As mentioned earlier, each PSs can feature various profiles starting from single-hump to double-hump and flat-top structure with different intermediate forms of symmetric and asymmetric nature. They result in an elastic interaction alone without any possibility for condensate density sharing, and one can refer \cite{tkpla14} for a detailed asymptotic analysis explaining the interaction. From the multi-profile nature of PSs, we can generate PSM by the known phase-matching condition, which leads to the bounded periodic attractive-repulsive wave pattern with a mixture of different profiles. For example, the molecule formed from a single-hump and double-hump PSs admits bounded propagation of three-hump oscillating breathers. Similarly, other possible PS structures also produce appropriate stationary or moving bound molecules. Further, the non-autonomous
	nonlinearity modulates the amplitude, width, and velocity of these PSM as explained previously for the FSM case. For
	completeness, we have depicted the elastic interaction of flat-top and
	double-hump (single-hump and asymmetric double-hump) PSs in $Q_2$
	($Q_1$ and $Q_3$) component(s) in the top row of Fig. \ref{fig-napsm}, where each PS experiences only a phase-shift
	after the interaction. However, their transition due to
	velocity-locking generates PSM with the multi-hump stationary structure
	for $k_{1I}=k_{2I}=0$ as shown in the second row of Fig. \ref{fig-napsm}, while this PSM can become a moving pattern for $k_{1I}=k_{2I}\neq 0$ (not shown
	here). The periodic and kink-like nonlinearity-induced multi-hump PSMs are also shown in the third and fourth rows of
	Fig. \ref{fig-napsm} with a transition from stationary to a left-moving molecule by
	tuning $\xi_2=-0.21$. One can also tune the $\alpha_j^{(u)}$ parameters in the other way to create PSM from diverse combinations of PSs.  
	
	\subsection{\bf Mixed Ferromagnetic-Polar Soliton Molecule (FPSM)} 
\begin{figure}
		\centering\includegraphics[width=0.27405\linewidth]{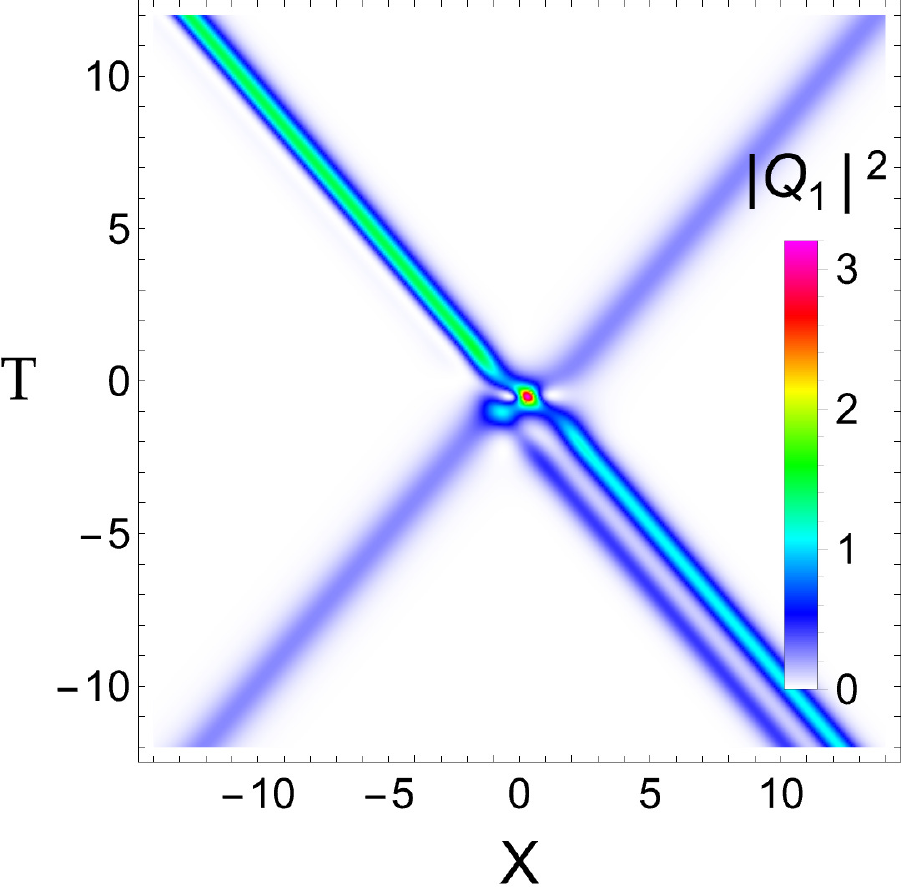}\quad \includegraphics[width=0.27405\linewidth]{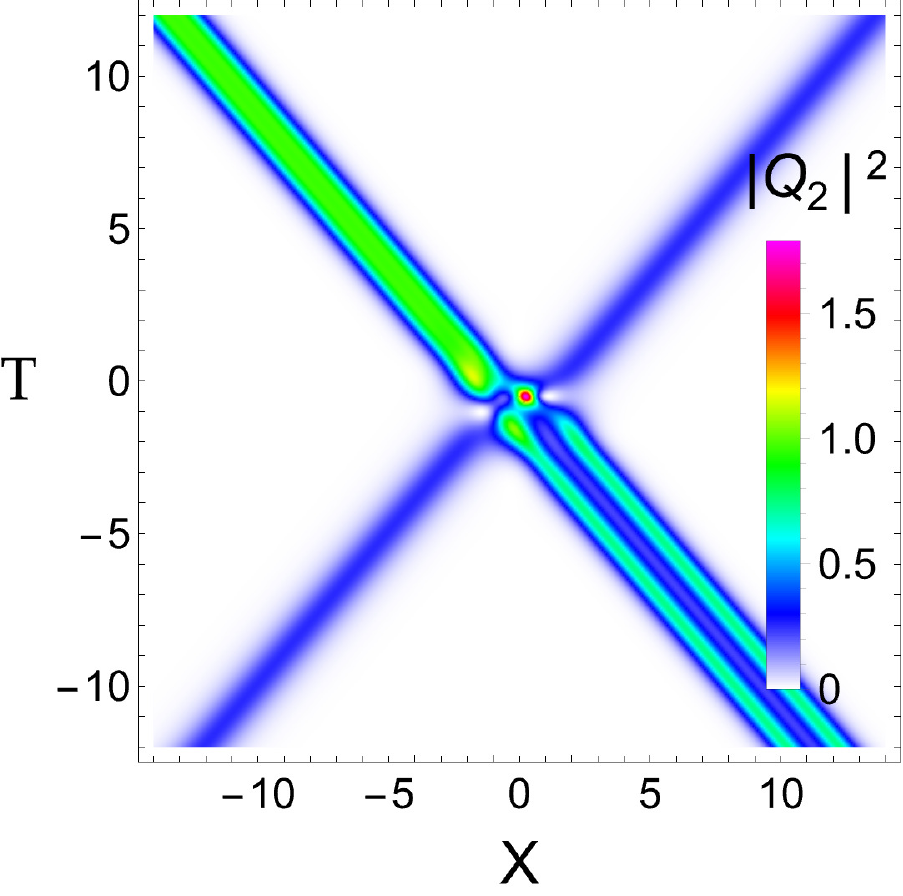}\quad
		\includegraphics[width=0.27405\linewidth]{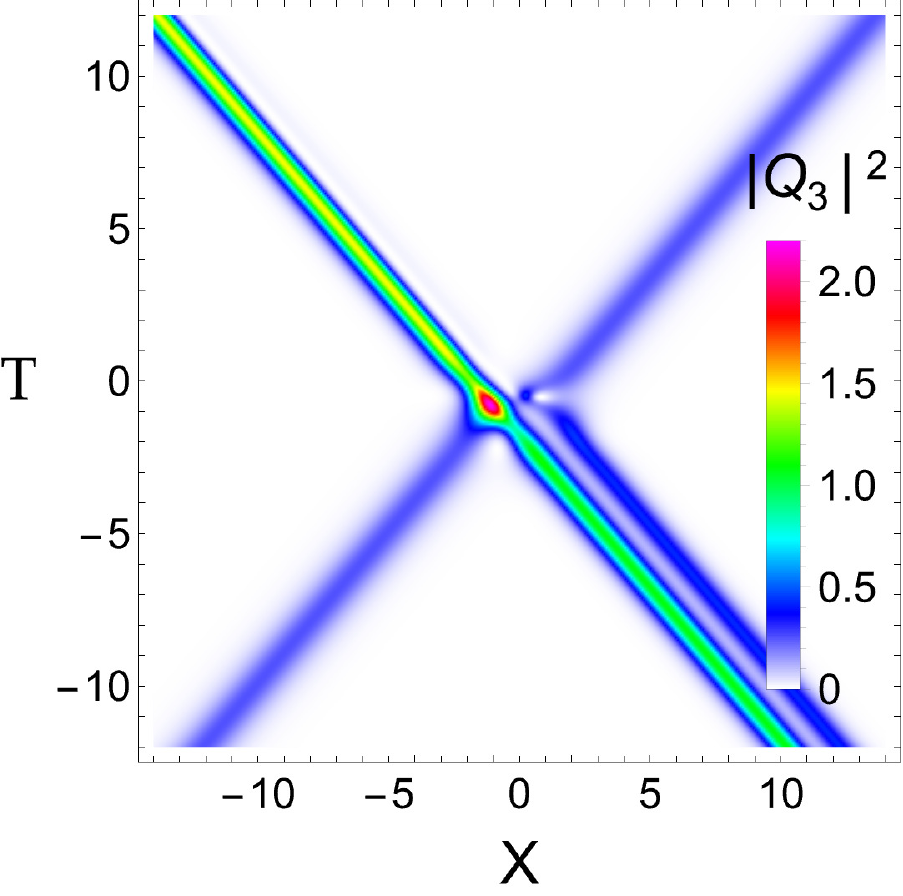}\\
		\centering\includegraphics[width=0.27405\linewidth]{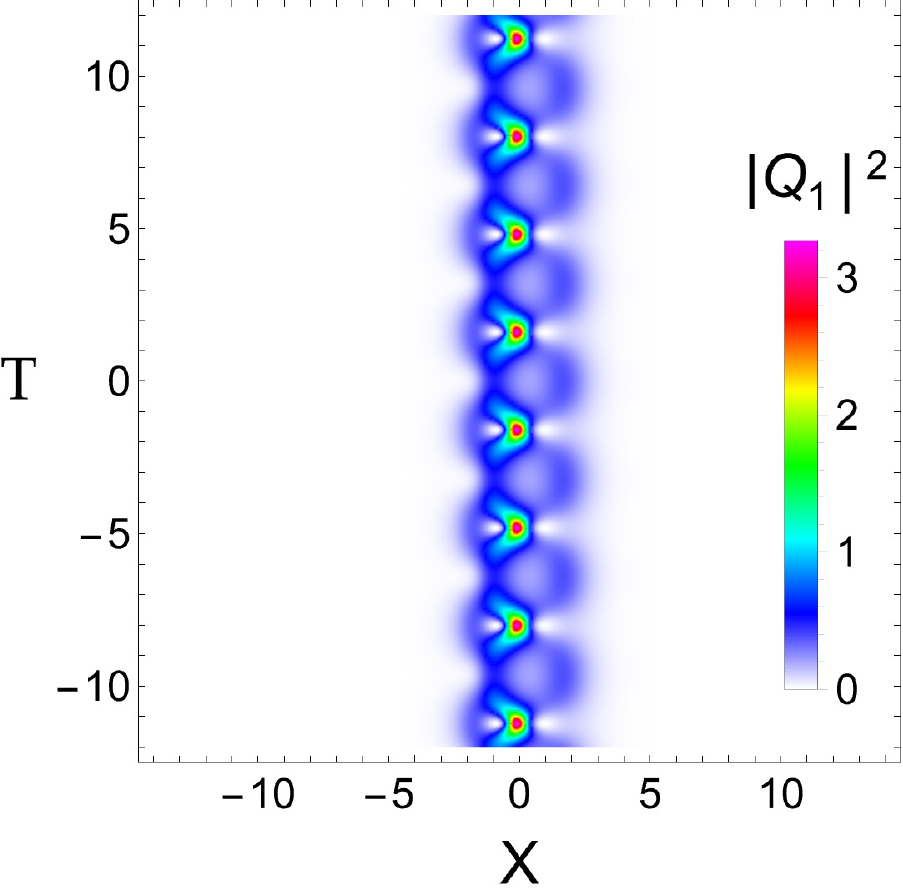}\quad \includegraphics[width=0.27405\linewidth]{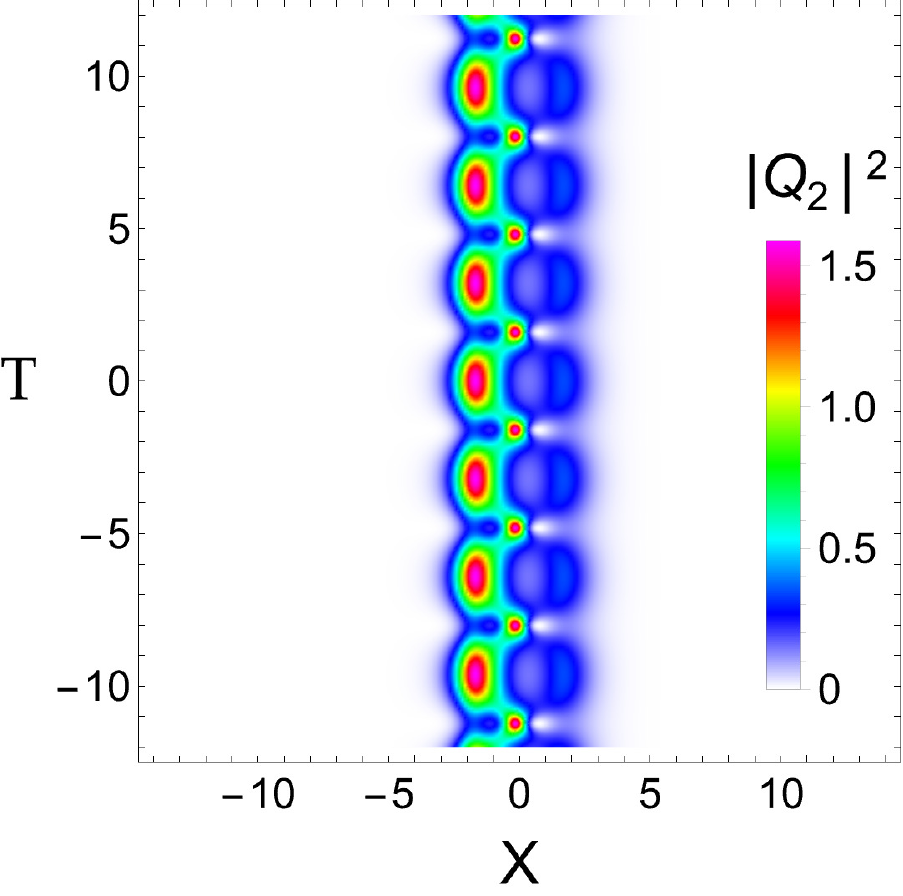}\quad
		\includegraphics[width=0.27405\linewidth]{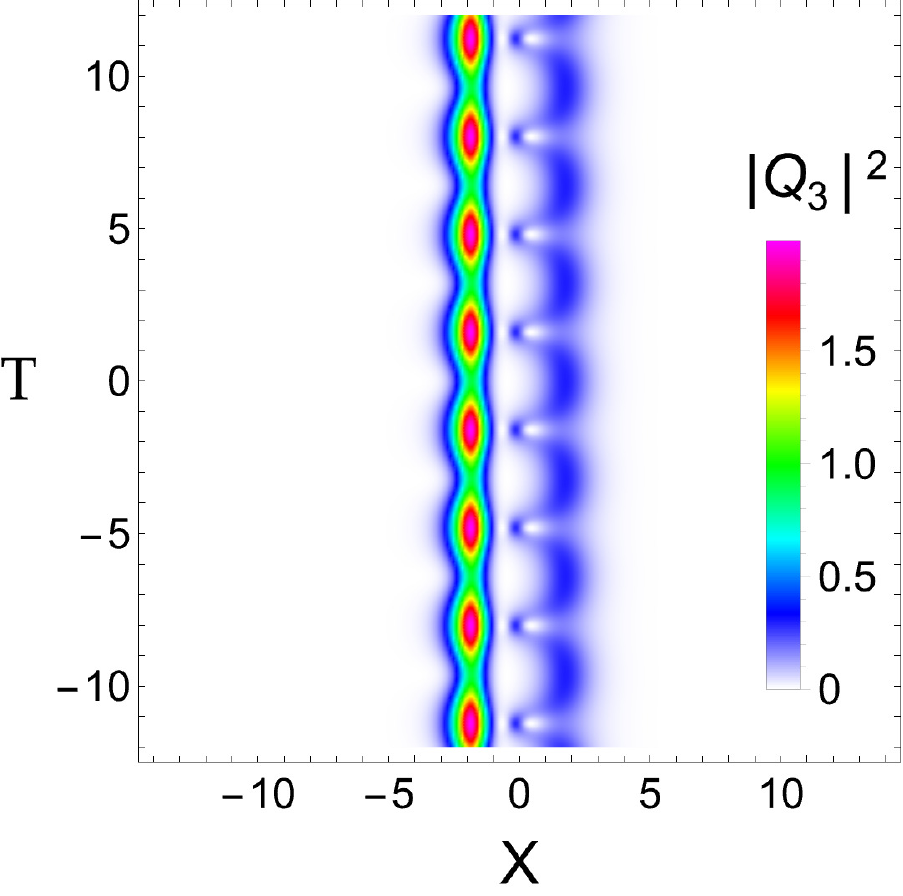}\\
		\centering\includegraphics[width=0.27405\linewidth]{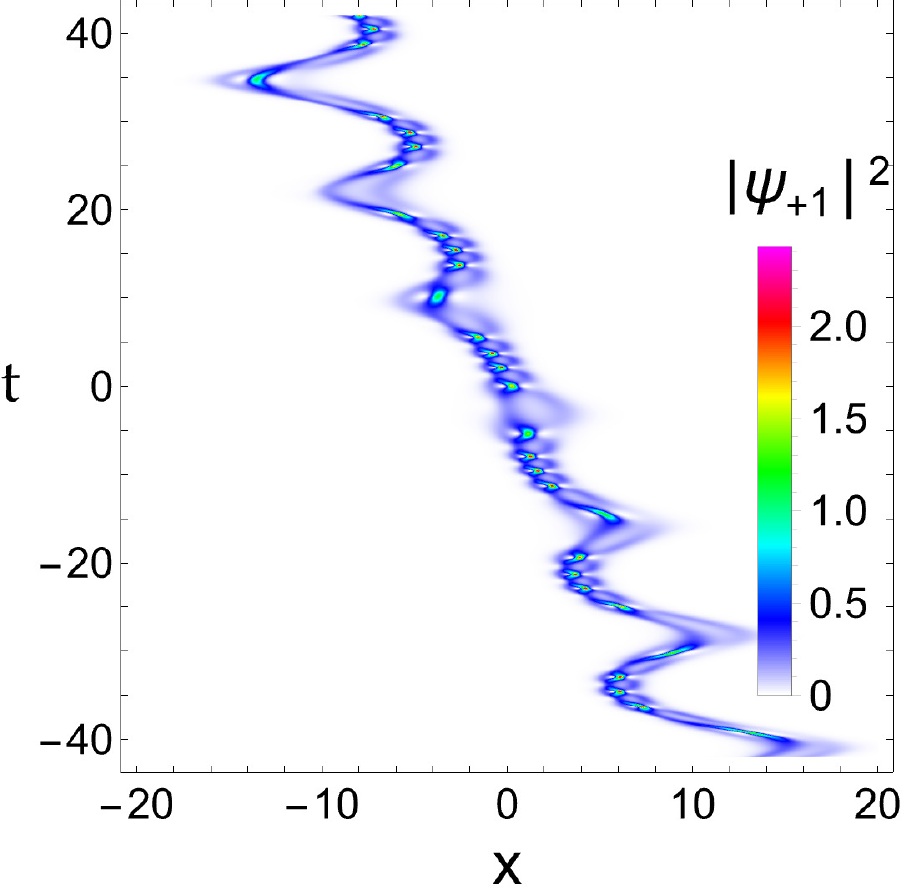}\quad \includegraphics[width=0.27405\linewidth]{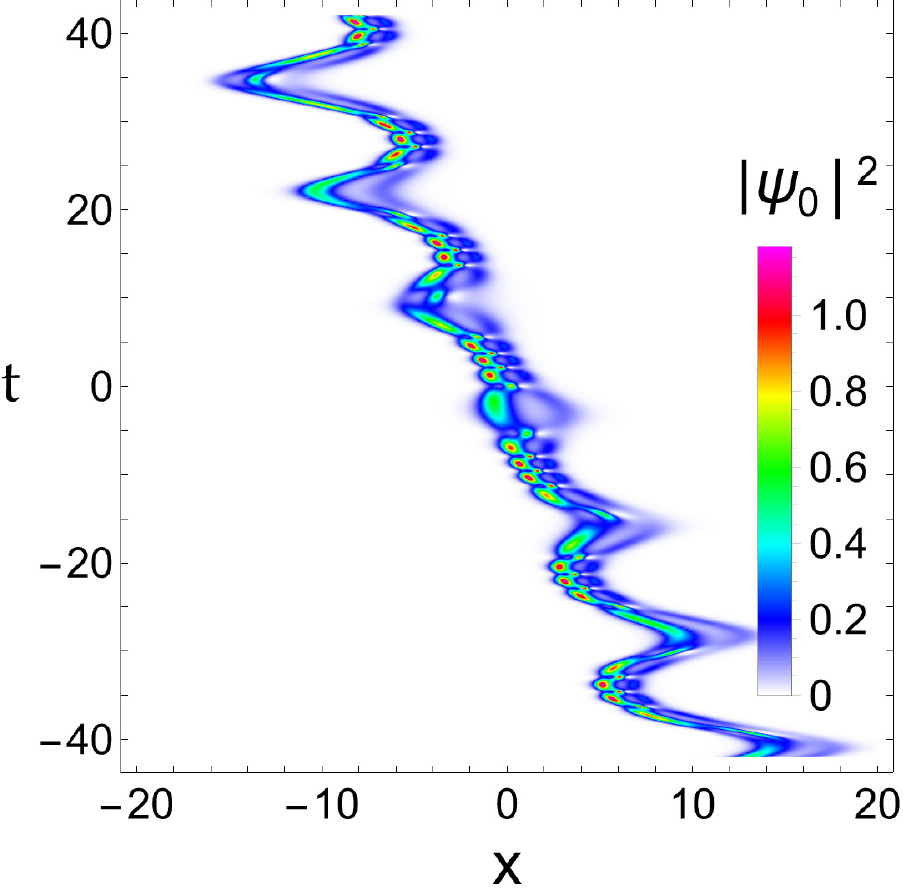}\quad 
		\includegraphics[width=0.27405\linewidth]{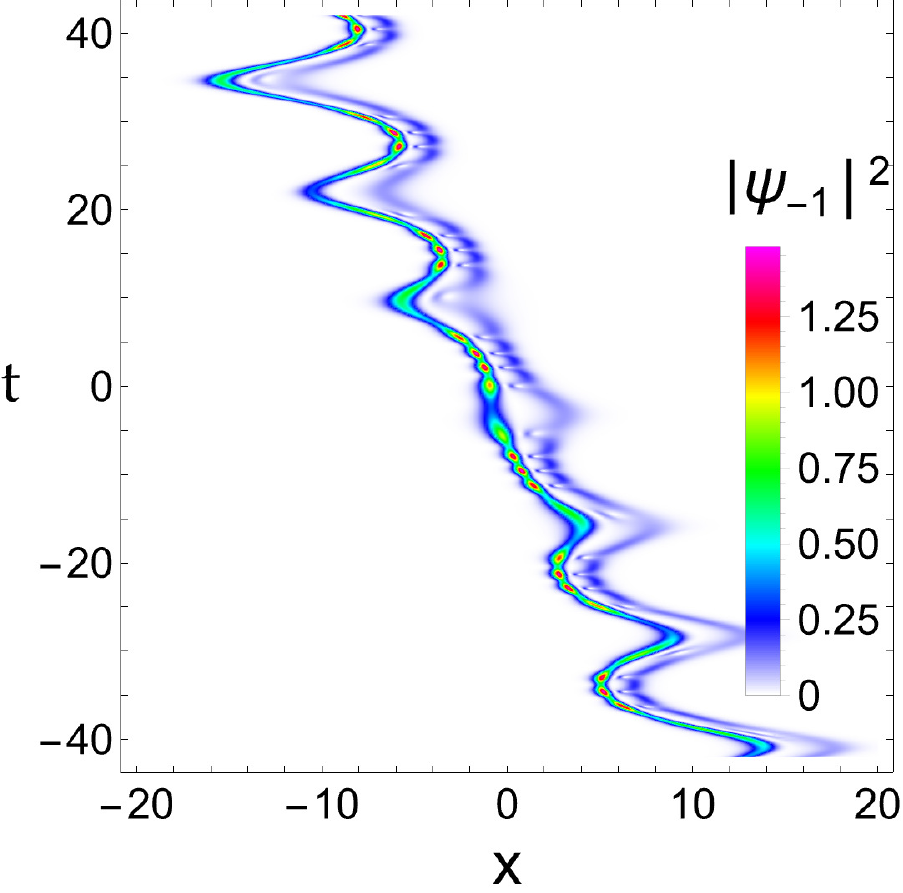}\\
		\centering\includegraphics[width=0.27405\linewidth]{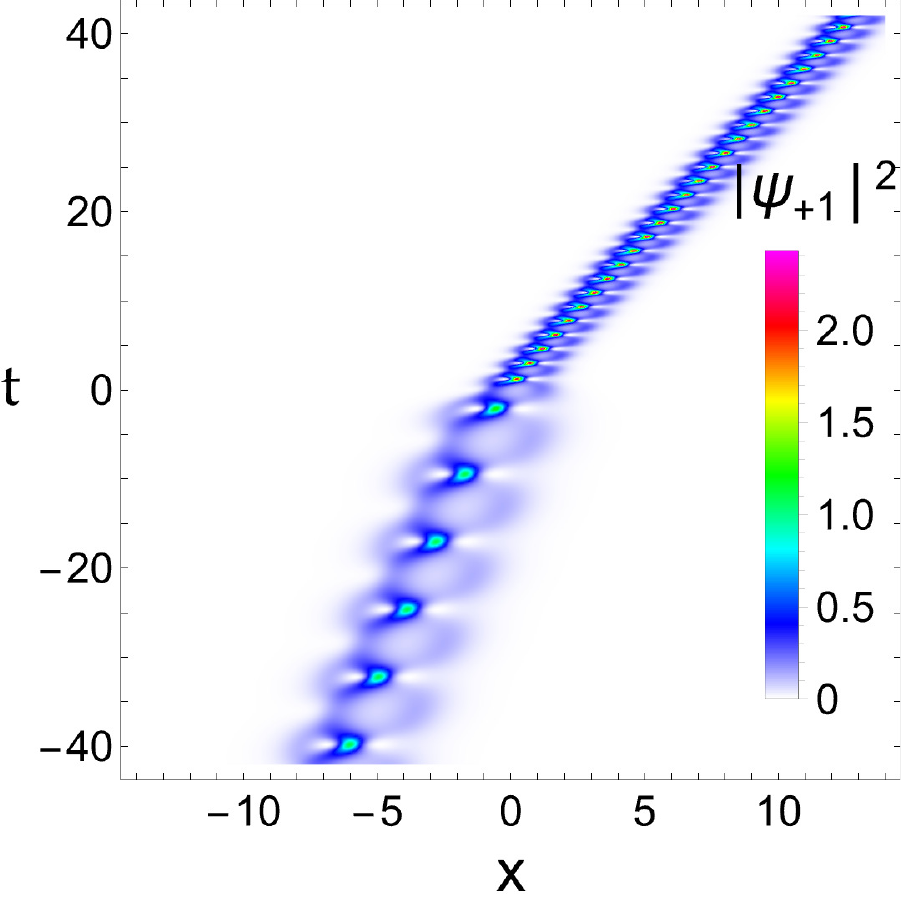}\quad \includegraphics[width=0.27405\linewidth]{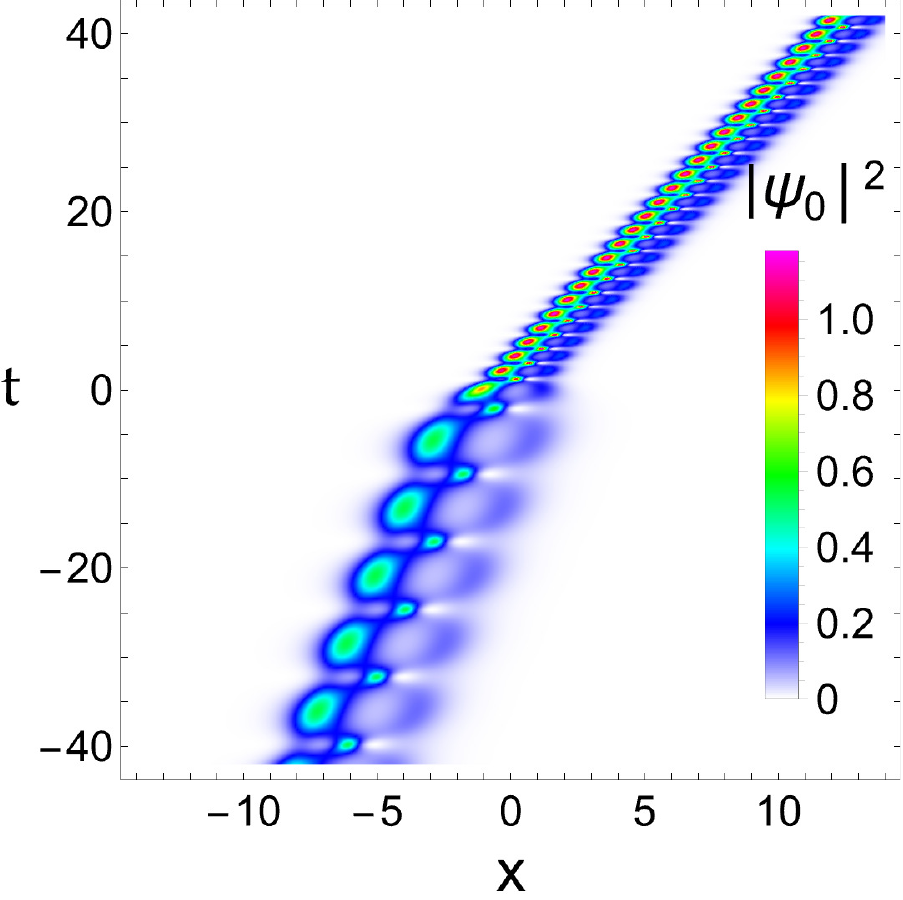}\quad 
		\includegraphics[width=0.27405\linewidth]{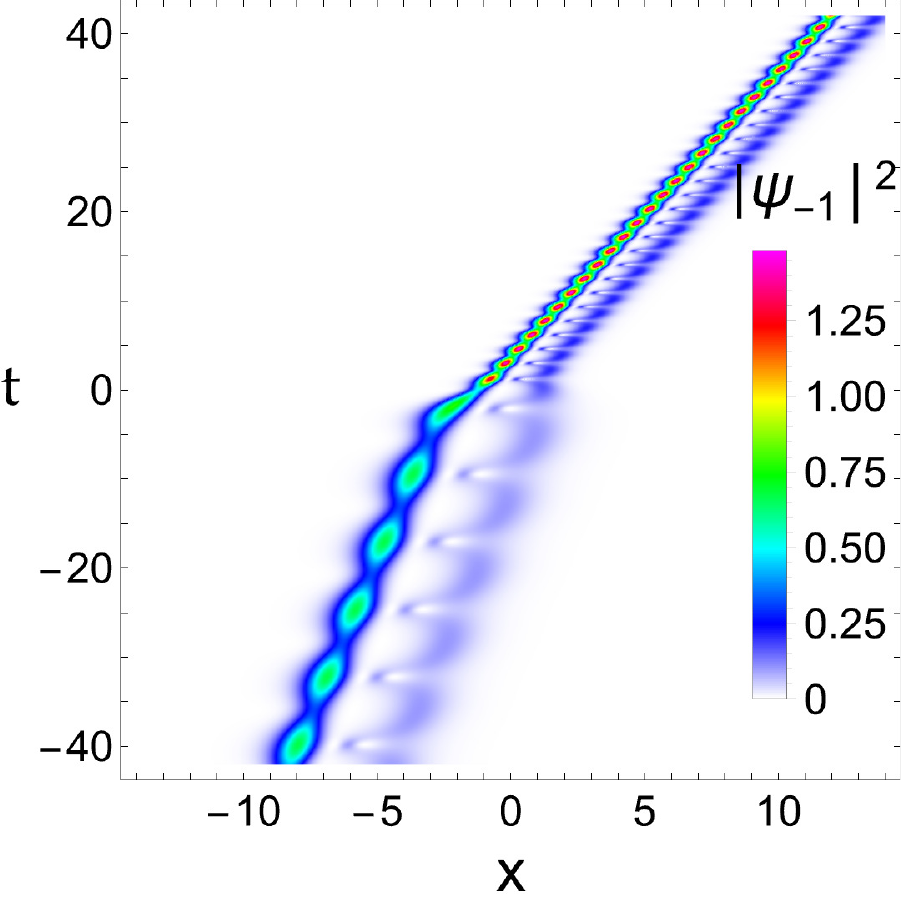}
		\caption{Spin-switching interaction of left-moving PS with right-moving FS for $\alpha_1^{(1)}=\alpha_1^{(2)}=\alpha_1^{(3)}=0.2$ and $\alpha_2^{(1)}=0.25,\alpha_2^{(2)}=-1.0,\alpha_2^{(3)}=0.75$. The PSs switch from the initial asymmetric/symmetric double-hump profile to single-hump/flat-top solitons and the FS reappears elastically (first row: $k_1=1+0.5i$ \& $k_2=1.7-0.5i$). The generation of stationary FPSM (second row: $k_1=1+0.0i$ \& $k_2=1.7+0.0i$) and their dynamics influenced by  non-autonomous periodic nonlinearity (third row: $\xi_1=0.52$ and $\xi_2=-0.21$) and kink-like nonlinearity (fourth row: $\xi_1=0.52$ and $\xi_2=0.21$).}
		\label{fig-nafpsm}
	\end{figure}

	The previous two cases deal with the dynamics of PSM and FSM generated from the pure state of two PSs or two FSs, respectively, where the former possesses the influence of spin-mixing nonlinearity and the latter does not have its role. The third branch of the two-soliton-pair can be constructed with a combination of both FS and PS, which can be attained either for $\Gamma_1= 0$ and $\Gamma_2\neq 0$ giving FS $S_1$ and PS $S_2$ or alternately $\Gamma_1\neq 0$ and $\Gamma_2= 0$ giving PS $S_1$ and FS $S_2$. However, for both possibilities, $\Gamma_3\neq 0$ and exclusively correlates the presence of spin-mixing nonlinearity. The mathematical expression for this case can be obtained from the general two-soliton solution (\ref{2sol}) for $\Gamma_1=0,~\Gamma_2\neq\Gamma_3\neq 0$ and 
	take the following simple exponential form:
	\bes\bea 
	G^{(j)}&=&\alpha _1^{(j)} e^{\eta _1} +\alpha _2^{(j)} e^{\eta _2} +e^{2\eta_2+\eta_1^*+\delta_{21}^{(j)}}+e^{2\eta_2+\eta_2^*+\delta_{22}^{(j)}}+e^{\eta_1+\eta_2+\eta_1^*+\delta_1^{(j)}}\nonumber\\
	&&+e^{\eta_1+\eta_2+\eta_2^*+\delta_2^{(j)}}+e^{\eta_1+2\eta_2+2\eta_2^*+\mu_{22}^{(j)}}+e^{\eta_1+\eta_1^*+2\eta_2+\eta_2^*+\mu_2^{(j)}},\\
	F&=&1+e^{\eta_{1}+\eta_{1}^*+R_1}+e^{\eta_{2}+\eta_{2}^*+R_2}+e^{\eta_1+\eta_2^*+\delta_0}+e^{\eta_2+\eta_1^*+\delta_0^*}+e^{2\eta_{2}+2\eta_{2}^*+\epsilon_{22}}\nonumber\\
	&&+e^{2\eta_{2}+\eta_1^*+\eta_{2}^*+\tau_2}+e^{\eta_{1}+\eta_2+2\eta_{2}^*+\tau_2^*}+e^{\eta_{1}+\eta_2+\eta_{1}^*+\eta_{2}^*+R_3}+e^{\eta_{1}+2\eta_2+\eta_{1}^*+2\eta_{2}^*+\theta_{22}}.\qquad 
	\eea 
	On simplification of the above $G^{(j)}$ and $F$, we can rewrite them in terms of trigonometric-hyperbolic functions as given below. 
	\bea 
	\hspace{-1.5cm}G^{(j)}_{FPSM}&=&e^{\frac{\delta_2^{(j)}+l_1^{(j)}}{2}}\cosh\left({\eta_{2R}}+N_{2R}^{(j)}+iN_{2I}^{(j)}\right)e^{i\eta_{1I}} + e^{\frac{\delta_1^{(j)}+l_2^{(j)}}{2}}\cosh\left({\eta_{1R}}+N_{1R}^{(j)}+iN_{1I}^{(j)}\right)e^{i\eta_{2I}},\nonumber\\
	&&+ e^{2\eta_{2R}+\frac{\mu_{22}+\delta_{21}^{(j)}}{2}}\cosh\left({\eta_{2R}}+N_{3R}^{(j)}+iN_{3I}^{(j)}\right)e^{i(\eta_{1I}+\eta_{2I})} \nonumber\\
	&&+ e^{2\eta_{2R}+\frac{\mu_2+\delta_{22}^{(j)}}{2}}\cosh\left({\eta_{1R}}+N_{4R}^{(j)}+iN_{4I}^{(j)}\right)e^{i\eta_{2I}},\quad j=1,2,3,\\
	\hspace{-1.5cm}F_{FPSM}&=& e^{\frac{\theta_{22}}{2}}\cosh\left({\eta_{1R}+2\eta_{2R}+{\theta_{22}}/{2}}\right) +e^{\frac{R_1+R_2}{2}}\cosh\left({\eta_{1R}-\eta_{2R}+{(R_1-R_2)}/{2}}\right)\nonumber\\
	&&+e^{2\eta_{2R}}\left(e^{\frac{R_3-\epsilon_{22}}{2}}\cosh\left({\eta_{1R}-\eta_{2R}+{(R_3+\epsilon_{22})}/{2}}\right) +e^{\tau_{2R}} \cos{(\eta_{2I}-\eta_{1I}+\tau_{2I})}\right) \nonumber\\
	&&+e^{\delta_{0R}} \cos{(\eta_{1I}-\eta_{2I}+\delta_{0I})},
	\eea 
	where $N_{1R}^{(j)}=\frac{\delta_{1R}^{(j)}-l_{2R}^{(j)}}{2}$, $N_{1I}^{(j)}=\frac{\delta_{1I}^{(j)}-l_{2I}^{(j)}}{2}$, $N_{2R}^{(j)}=\frac{\delta_{2R}^{(j)}-l_{1R}^{(j)}}{2}$, $N_{2I}^{(j)}=\frac{\delta_{2I}^{(j)}-l_{1I}^{(j)}}{2}$, $N_{3R}^{(j)}=\frac{\mu_{22R}^{(j)}-\delta_{21R}^{(j)}}{2}$, $N_{3I}^{(j)}=\frac{\mu_{22I}^{(j)}-\delta_{21I}^{(j)}}{2}+\eta_{1I}-\eta_{2I}$,  $N_{4R}^{(j)}=\frac{\mu_{2R}^{(j)}-\delta_{22R}^{(j)}}{2}$, and $N_{4I}^{(j)}=\frac{\mu_{2I}^{(j)}-\delta_{22I}^{(j)}}{2}$.
	Thus, the resulting explicit form of autonomous and non-autonomous FPSMs are expressed as 
	\bea 
	&& Q_j(X,T) = \frac{G^{(j)}_{FPSM}(X,T)}{F_{FPSM}(X,T)}, \qquad j=1,2,3,\label{afpsm}\\
	&&\left(\psi_{+1},~ \psi_{0},~\psi_{-1}\right)^T = \xi_1\sqrt{{c}(t)}~ (Q_1,~Q_2,~Q_3)^T e^{i\theta(x,t)},\qquad j=1,2,3.\qquad\qquad\label{nafpsm}
	\eea \ees 
	A careful analysis of the interaction behaviour shows that the PS undergoes inelastic (spin-mixing) interaction leaving the FS unaltered along all the components, and both solitons exhibit an additional phase-shift. Particularly, during such spin-mixing interaction, PS can change its profile nature from single-hump/flat-top to symmetric/asymmetric double-hump or vice-versa, which can be tuned by using the $\alpha_u^{(j)}$ parameters. Detailed asymptotic analysis describing such FS$\times$PS interaction is provided in Ref. \cite{tkpla14}. Now, we focus on the mixed FPSM obtained for the already identified choice of velocity resonance $k_{2I}=k_{1I}$ and thereby the resulting FPSM provides stationary or travelling patterns possessing various combinations of profile structures based on the selection of $\alpha_u^{(j)}$, $k_{1R}$, $k_{2R}$ and $k_{1I}$ parameters. 
	We have shown the spin-switching interaction of PS ($S_2$) with FS ($S_1$) in the top panels of Fig.~\ref{fig-nafpsm}. The left-moving PS ($S_2$) has an asymmetric (symmetric) double-hump profile in $Q_{1,3}$ ($Q_2$) components and switches to a single-hump (flat-top) profile with suppression (enhancement) appropriately conserving the total density of molecules. However, as predicted, the right-moving FS ($S_1$) undergoes elastic interaction in all three components. The second panel shows the stationary FPSM and the temporal dependence of nonlinearity $c(t)$ does not break the molecule similar to the previous cases. During the molecule formation, these solitons do not pass through each other, except for periodic head-on (touching or kissing) interactions.
	Additionally, the dynamics of these FPSMs can also be manipulated accordingly using the presence of arbitrary temporally-varying nonlinearities $c(t)$ and similarity parameters $\xi_1$ and $\xi_2$. The nature of the bound FPSM structure gets altered according to the periodic and kink-like nonlinearities as given in the previous two cases with different profile patterns. For completeness, we have shown the corresponding modulated FPSMs in the third and fourth panels of Fig. \ref{fig-nafpsm}. One can observe these switching interactions in several other possibilities by suitably playing with the spin-polarization parameters $\alpha_u^{(j)}$, $u=1,2,~j=1,2,3$.
	
	Considering the length of the manuscript, we have investigated only two types of temporally varying nonlinearity modulations (periodic and kink-like variation driven by snoidal function). One can extend the analysis further by explaining with respect to various other types of modulations and corresponding dynamics of soliton molecules. The profile-preserving nature of these stationary spinor bright bound SMs indicates their stable propagation characteristics. However, its dynamics under interaction with another soliton or molecule require a dedicated analysis and can be considered for future investigation. Additionally, exploring the stability of SMs using numerical analysis will also be of an immediate objective. Beyond these simple matter-wave soliton molecules (FSM, PSM and FPSM) containing only two solitons, it is of our natural interest to extend the analysis by incorporating multiple bright, dark, and mixed bright-dark solitons as well as their evolution under autonomous and non-autonomous settings. 
	
	\section{Conclusions}\label{sec-conclu} 
	In this work, we have considered a set of three-coupled Gross-Pitaevskii equations describing the dynamics of spinor $F=1$ Bose-Einstein condensates with non-autonomous nonlinearities that can be controlled by the Feshbach mechanism. By using the exact soliton solutions, we have studied the possibility of generating matter-wave bright soliton molecules (SMs) through the velocity resonance mechanism and explored their dynamics in detail. Especially the obtained SMs are classified based on the nature of participating soliton as ferromagnetic soliton molecules (FSM), polar soliton molecules (PSM), and mixed ferromagnetic-polar soliton molecules (FPSM). Their characteristics and the role of time-varying nonlinearities in each of these classes of molecules are investigated. These bound soliton molecules modulate themselves during propagation by keeping their stable properties. Particularly, the modulation of spinor SMs under periodic and kink-like nonlinearities revealed the snaking and single-step compressed amplification of multi-structured breathing patterns along with appreciable changes in their amplitude, velocity, width and period of oscillations without breaking the bound structure. The presented results are significant as they go beyond the known interaction dynamics of matter-wave bright solitons in spinor condensates. In particular, the interesting soliton molecule bending can have immediate applications in atomic/optical switching devices and their optical counterpart in communication. The reason for this is without the loss of energy, the track of the optical pulse can be altered and even it can be amplified through the energy-sharing process among the different modes in the case of multimode propagation. Beyond the above, one can focus to study the dynamics of dark-dark and bright-dark multi-soliton molecules in spinor BECs and the interaction of molecules with individual solitons as an immediate future assignment to get more impetus into the influence of inhomogeneities in the system. Such studies will have ramifications in multimode graded-index fibers in addition to atomic condensates.\\
	
	\setstretch{1.075}
	\noindent{\bf Acknowledgement}\\
	The research work of K Sakkaravarthi is supported by the Young Scientist Training (YST) program of the Asia-Pacific Center for Theoretical Physics (APCTP), Pohang-si, Gyeongsangbuk-do, Republic of Korea. The APCTP is supported by the Korean Government through the Science and Technology Promotion Fund and Lottery Fund. T Kanna acknowledges the support received from the Department of Science and Technology - Science and Engineering Research Board (DST-SERB), Government of India, through a Core Research Grant No. CRG/2021/004119. {The authors thank the handling editor and anonymous reviewers for providing valuable comments and suggestions that helped immensely to improve the quality of the manuscript.}\\ 
	
	\noindent{\bf CRediT Authorship Contribution Statement}\newline {\bf K. Sakkaravarthi}: Conceptualization, Formal Analysis and Investigation, Validation, Resources, Writing - Original Draft Preparation, Writing - Review \& Editing.\newline {\bf R. Babu Mareeswaran}: Methodology, Writing - Review \& Editing. \newline 
	{\bf T. Kanna}: Methodology, Resources, Writing - Review \& Editing, Supervision.\\~\\
	
	\noindent{\bf Declaration of Competing Interest}\\ The authors declare that they have no known competing financial interests or personal relationships that could have appeared to influence the work reported in this paper.
	
	\appendix
	\section{Bright matter-wave two-soliton solution}\label{appendix}
	The explicit form of functions appearing in the bright two-soliton solution (\ref{2sol}) can be expressed as below \cite{tkwcna,tkpla14}.
	\bes\bea
	\hspace{-1.95cm}G^{(j)}&=&\alpha _1^{(j)} e^{\eta _1} +\alpha _2^{(j)} e^{\eta _2}+\sum e^{2 \eta _u+\eta _v^{*}+\delta_{uv}^{(j)}}+\sum e^{\eta_1+\eta_2+\eta_u^*+\delta_u^{(j)}}+\sum e^{2\eta_u+2\eta_v^*+\eta_{3-u}+\mu_{uv}^{(j)}} \nonumber\\
	\hspace{-1.95cm}&&+e^{\eta_1+\eta_1^*+\eta_2+\eta_2^*}\left(\sum e^{\eta_u+\mu_u^{(j)}}+\sum e^{\eta_1+\eta_2+\eta_u^*+\phi_u^{(j)}}\right),\quad j=1,2,3,\\
	\hspace{-1.95cm}F&=&1+\sum e^{\eta_u+\eta_u^*+R_u}+e^{\eta_1+\eta_2^*+\delta_0}+e^{\eta_2+\eta_1^*+\delta_0^*}+\sum e^{2\eta_u+2\eta_v^*+\epsilon_{uv}}+e^{\eta_1^*+\eta_2^*}\sum e^{2\eta_u+\tau_u}\nonumber\\
	\hspace{-1.95cm}&& +e^{\eta_1+\eta_2}\sum e^{2\eta_u^*+\tau_u^*}+e^{\eta_1+\eta_1^*+\eta_2+\eta_2^*}\left(e^{R_3}+\sum e^{\eta_u+\eta_v^*+\theta_{uv}}+e^{\eta_1+\eta_1^*+\eta_2+\eta_2^*+R_4}\right),~~~~~
	\eea
	and the auxiliary function $S$ becomes
	\bea
	\hspace{-2.5cm}S=\sum \Gamma_u e^{2 \eta _u}+\Gamma_3 e^{\eta _1+\eta _2} +\sum e^{\eta _u+2\eta _{3-u}+\eta _v^*+\lambda _{uv}} + e^{2 \eta _1+2 \eta _2} \left(\sum e^{2\eta _u^*+\lambda _u}+e^{\eta_1^*+\eta _2^*+\lambda _3}\right),\qquad 
	\eea
	where $\eta_u=k_u(X+ik_uT)$ and every summation is taken over $u=1,2,$ and/or $v=1,2$, with other parameters as listed below. 
	\bea
	\hspace{-2.5cm}e^{R_u}&=&\frac{\kappa _{uu}}{(k_u+k_u^*)},~ e^{\delta _0}=\frac{ \kappa _{12}}{(k_1+k_2^*)},~
	e^{\delta _0^*}=\frac{ \kappa _{21}}{(k_2+k_1^*)},~
	e^{\delta _{uv}^{(j)}}=\frac{(-1)^{j+1} \alpha_v^{(4-j)*} \Gamma_u}{(k_u+k_v^*)^2},~~\\
	\hspace{-2.5cm}e^{\delta _u^{(j)}}&=&\frac{(-1)^{j+1} \alpha _u^{(4-j)*} \Gamma_3+(k_1-k_2) (\alpha _1^{(j)} \kappa _{2u}-\alpha _2^{(j)} \kappa_{1u})}{(k_1+k_u^*) (k_2+k_u^*)},~~
	e^{\epsilon _{uv}}=\frac{\Gamma_u \Gamma_v}{(k_u+k_v^*)^4},\\ 
	\hspace{-2.5cm}e^{\tau _u}&=&\frac{\Gamma_u \Gamma_3}{ (k_u+k_1^*)^2 (k_u+k_2^*)^2},~~~~
	e^{\theta _{uv}}=\frac{|k_1-k_2|^4 }{{\tilde{D}}(k_u+k_v^*)^2} \Gamma_u \Gamma_v^* \kappa_{3-u3-v},\\
	\hspace{-2.5cm}e^{\lambda _{uv}}&=&\frac{(k_1-k_2)^2 \kappa_{uv} \Gamma_{3-u}}{(k_u+k_v^*)(k_{3-u}+k_v^*)^2},~~~~
	e^{\mu _{uv}^{(j)}}=\frac{(k_1-k_2)^2 \alpha _{3-u}^{(j)} \Gamma_u \Gamma_v^*}{(k_u+k_v^*)^4 (k_{3-u}+k_v^*)^2},~\\
	\hspace{-2.5cm}e^{\lambda _u}&=&\frac{(k_1-k_2)^4 ~\Gamma_1 \Gamma_2 \Gamma_u^*}{(k_1+k_u^*)^4 (k_2+k_u^*)^4},~~~~~~~
	e^{\lambda _3}=\frac{(k_1-k_2)^4 \Gamma_1 \Gamma_2 \Gamma_3}{\tilde{D}},\\
	\hspace{-2.5cm}e^{\phi_u^{(j)}}&=&(-1)^{(j+1)}{\alpha _{3-u}^{(4-j)*}}\frac{(k_1-k_2)^4 (k_1^*-k_2^*)^2~ \Gamma_1 \Gamma_2 \Gamma_u^*} {\tilde{D} {(k_1+k_u^*)^2(k_2+k_u^*)^2}},\\
	\hspace{-2.5cm}e^{R_3}&=& \frac{|k_1-k_2|^2(\kappa_{11}\kappa_{22}-\kappa_{12}\kappa_{21})+|\Gamma_3|^2}{(k_1+k_1^*)|k_1+k_2^*|^2(k_2+k_2^*)}, ~~
	e^{R_4}=\frac{|k_1-k_2|^8 |\Gamma_1|^2 |\Gamma_2|^2}{\tilde{D}^2},~~~~~\\
	\hspace{-2.5cm}e^{\mu_u^{(j)}}&=&\frac{(k_1-k_2)^2~\Gamma_u \left(\alpha_{3-u}^{(j)}\Gamma_3^*+(-1)^{(j+1)}(k_1^*-k_2^*)(\alpha_1^{(4-j)*}\kappa_{3-u2}-\alpha_2^{(4-j)*}\kappa_{3-u1})\right)}{\sqrt{\tilde{D}}(k_u+k_1^*)(k_u+k_2^*)},~~~\\
	\hspace{-2.5cm}\Gamma_u&=& \alpha_u^{(1)}\alpha_u^{(3)}-(\alpha_u^{(2)})^2,\qquad  \Gamma_3= \alpha_1^{(1)}\alpha_2^{(3)}+\alpha_2^{(1)}\alpha_1^{(3)}-2\alpha_1^{(2)}\alpha_2^{(2)},\\
	\hspace{-2.5cm}\tilde{D}&=&(k_1+k_1^*)^2(k_1^*+k_2)^2 (k_1+k_2^*)^2 (k_2+k_2^*)^2,\\
	\hspace{-2.5cm}\kappa_{uv}&=&\frac{ \alpha_u^{(1)} \alpha_v^{(1)*}+2\alpha_u^{(2)} \alpha_v^{(2)*}+\alpha_u^{(3)} \alpha_v^{(3)*}}{k_u+k_v^*}.
	\eea\ees
	In the above equations $u,v=1,2$, and $j=1,2,3$.
	
	\setstretch{01.20}
	
\end{document}